\documentclass[prx,aps,twocolumn,floatfix,nofootinbib,preprintnumbers,superscriptaddress,amsmath,amssymb,longbibliography]{revtex4-1}

\usepackage{changes}
\definechangesauthor[name={Xiaojian Bai}, color=purple]{XB}

\usepackage[%
  colorlinks=true,
  urlcolor=blue,
  linkcolor=blue,
  citecolor=blue
]{hyperref}

\usepackage{xcolor,bm,lineno,multirow,float,times}
\usepackage{physics}
\usepackage{rotating}
\usepackage{verbatim}
\usepackage{graphicx}
\usepackage[multiple]{footmisc}
\usepackage{natbib}
\usepackage{soul}

\def\Ang{\AA$^{-1}$ }

\newcommand\blfootnote[1]{%
	\begingroup
	\renewcommand\thefootnote{}\footnote{#1}%
	\addtocounter{footnote}{-1}%
	\endgroup}

\begin{document}


\title{Quantum versus Classical Spin Fragmentation in Dipolar Kagome Ice Ho${_3}$Mg${_2}$Sb${_3}$O${_{14}}$}

\author{Zhiling Dun}
\email{zdun3@gatech.edu}
\affiliation{School of Physics, Georgia Institute of Technology, Atlanta, GA 30332, USA}
\affiliation{Department of Physics and Astronomy, University of Tennessee, Knoxville, TN 37996, USA}

\author{Xiaojian Bai}
\blfootnote{Z. Dun and X. Bai contributed equally to this work.}
\affiliation{School of Physics, Georgia Institute of Technology, Atlanta, GA 30332, USA}
	
\author{Joseph A.~M. Paddison}
\email{paddisonja@ornl.gov}
\affiliation{School of Physics, Georgia Institute of Technology, Atlanta, GA 30332, USA}
\affiliation{Churchill College, University of Cambridge, Storey's Way, Cambridge CB3 0DS, United Kingdom}
\affiliation{Materials Science and Technology Division, Oak Ridge National Laboratory, Oak Ridge, TN 37831, USA}

\author{Emily Hollingworth}
\affiliation{School of Physics, Georgia Institute of Technology, Atlanta, GA 30332, USA}

\author{Nicholas P. Butch}
\affiliation{NIST Center for Neutron Research,  Gaithersburg, MD 20899, USA}

\author{Clarina D. Cruz}
\affiliation{Neutron Scattering Division, Oak Ridge National Laboratory, Oak Ridge, TN 37831, USA}

\author{Matthew B. Stone}
\affiliation{Neutron Scattering Division, Oak Ridge National Laboratory, Oak Ridge, TN 37831, USA}

\author{Tao Hong}
\affiliation{Neutron Scattering Division, Oak Ridge National Laboratory, Oak Ridge, TN 37831, USA}

\author{Franz Demmel}
\affiliation{ISIS Facility, Rutherford Appleton Laboratory, Didcot, OX11 0QX, United Kingdom}

\author{Martin Mourigal}
\email{mourigal@gatech.edu}
\affiliation{School of Physics, Georgia Institute of Technology, Atlanta, GA 30332, USA}

\author{Haidong Zhou}
\email{hzhou10@utk.edu}
\affiliation{Department of Physics and Astronomy, University of Tennessee, Knoxville, TN 37996, USA}
\affiliation{National High Magnetic Field Laboratory, Florida State University, Tallahassee, FL 32310, USA}

\date{\today}

\begin{abstract}
A promising route to realize entangled magnetic states combines geometrical frustration with quantum-tunneling effects. Spin-ice materials are canonical examples of frustration, and Ising spins in a transverse magnetic field are the simplest many-body model of quantum tunneling. Here, we show that the tripod kagome lattice material Ho${_3}$Mg${_2}$Sb${_3}$O${_{14}}$  unites an ice-like magnetic degeneracy with quantum-tunneling terms generated by an intrinsic splitting of the Ho$^{3+}$ ground-state doublet, which is further coupled to a nuclear spin bath. Using neutron scattering and thermodynamic experiments, we observe a symmetry-breaking transition at $T^{\ast}\approx0.32$\,K to a remarkable state with three peculiarities: a concurrent recovery of magnetic entropy associated with the strongly coupled electronic and nuclear degrees of freedom; a fragmentation of the spin into periodic and ice-like components; and persistent inelastic magnetic excitations down to $T\approx0.12$\,K. These observations deviate from expectations of classical spin fragmentation  on a kagome lattice, but can be understood within a model of dipolar kagome ice under a homogeneous transverse magnetic field, which we survey with exact diagonalization on small clusters and mean-field calculations. In Ho${_3}$Mg${_2}$Sb${_3}$O${_{14}}$, hyperfine interactions dramatically alter the single-ion and collective properties, and suppress possible quantum correlations, rendering the fragmentation with predominantly single-ion quantum fluctuations. Our results highlight the crucial role played by hyperfine interactions in frustrated quantum magnets, and motivate further investigations of the role of quantum fluctuations on partially-ordered magnetic states.

\end{abstract} 

\maketitle

\section{Introduction}
Quantum spin liquids are exotic states of magnetic matter in which conventional magnetic order is suppressed by strong quantum fluctuations \cite{Balents_2010}. Frustrated magnetic materials, which have a large degeneracy of classical magnetic ground states, are often good candidates to search for this elusive behavior. A canonical example of frustration is spin ice, in which Ising spins occupy a pyrochlore lattice of corner-sharing tetrahedra \cite{Harris_1997,Bramwell_2001}. Classical ground states obey the ``two in, two out" ice rule for spins on each tetrahedron, and thermal excitations behave as deconfined magnetic monopoles~\cite{Castelnovo_2008,morris2009dirac, kadowaki2009observation, Fennell_2009}. These pairs of fractionalized excitations interact \textit{via} Coulomb's law and correspond to topological defects of a classical field theory obtained by coarse-graining spins into a continuous magnetization. In principle, topological \emph{quantum} excitations can be generated by adding quantum-tunneling terms to the classical spin-ice model---e.g., by adding couplings between the transverse components of spins~\cite{Hermele_2004,Savary_2012,Gingras_2014}, or by introducing a local magnetic field transverse to the Ising spins \cite{Moessner_2000,Henry_2014,Tomasello_2015,Savary_2017}. A search for real materials that realize such quantum spin-ice states has found several promising candidates (see, e.g., \cite{Zhou_2008,Ross_2011,Thompson_2011,Fennell_2012,Sibille_2015,Sibille_2016,Petit_2016,Wen_2017,lhotel2018evidence,Sibille_2018,Mauws_2018}). However, important challenges remain, including the determination of the often-complex spin Hamiltonian \cite{Jaubert_2015,Yan_2017,Thompson_2017}, the subtle role that structural disorder may play \cite{Sala_2014,Martin_2017,Mostaed_2017}, and the computational challenges associated with simulations of three-dimensional (3D) quantum magnets \cite{Shannon_2012,Kato_2015}.

\begin{figure}[tbp] 
	\begin{center}
		\includegraphics[width=\columnwidth]{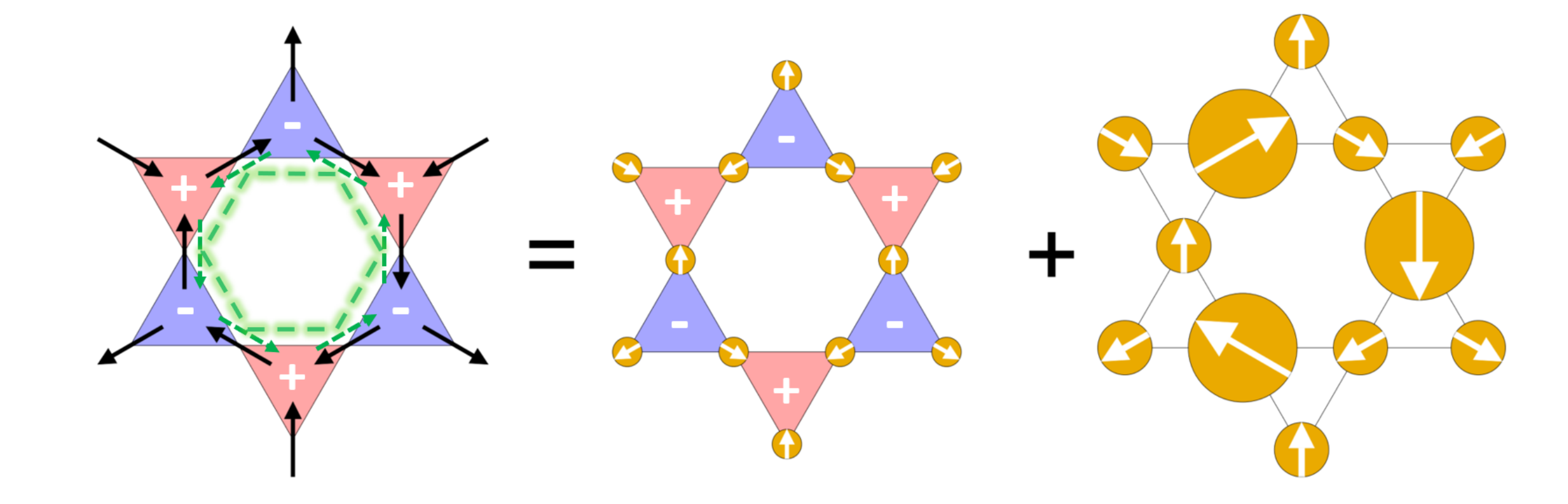}
	\end{center}
	\caption{\label{fig1}  Classical spin fragmentation (CSF) process in a model of dipolar kagome ice displaying emergent-charge order, based on Ref.~\onlinecite{Canals_2016}. The expectation values of spins $\langle\sigma^z_i\rangle$ are represented by black arrows. Each triangle has one spin pointing ``in" (towards its center) and two pointing ``out" (away from its center), or \emph{vice versa}. The emergent magnetic charge of a triangle is defined as the number of spins pointing ``in" minus the number pointing ``out". Positive ($Q_j=+$) and negative ($Q_j=-$) emergent charges are represented as red and blue triangles, respectively, and form a staggered arrangement. Three distinct spin configurations are possible for a given emergent charge, which yields a macroscopic number of degenerate spin configurations associated with emergent charge ordering. Spin fragmentation decomposes each unit-length spin into ``divergence-full'' and ``divergence-free'' channels (center and right images, respectively). The fragmented spins are shown as orange circles with diameter proportional to the length of the fragmented spin. The green hexagon represents the flipping of six spins around a closed loop: this is the simplest process that connects two distinct spin configurations within the degenerate CSF manifold.}
\end{figure}

A promising alternative route towards quantum analogs of spin ice is offered by two-dimensional Ising ferromagnets on a kagome lattice. When the spins are confined to point either towards or away from the center of each triangle of the lattice, a highly degenerate kagome ice state is stabilized with a ``one in, two out" or ``two in, one out" local ice rule on each triangle~\cite{Wills_2002}. Both a quantum-tunneling term and an external magnetic field are required to enable tunneling between these states \cite{Carrasquilla_2015,Wu_2018}. Remarkably, when the long range magnetic dipole-dipole interaction is introduced, the effective Coulomb interaction between emergent magnetic charges---defined in Fig.~\ref{fig1}---selects a sub-space of the kagome ice manifold and drives a phase transition to an intermediate-temperature phase with staggered emergent-charge ordering \cite{Moller_2009,Chern_2011}. This state possesses nonzero entropy because each emergent charge retains a threefold degeneracy of spin orientations \cite{Moller_2009}; hence, ordering of the emergent  charges does not imply complete long-range ordering of the spins. Fig.~\ref{fig1} shows that such spin structures can be decomposed into a ``divergence-full'' channel in which spins are spatially ordered, and an ``divergence-free'' channel in which spins remain spatially disordered---a process known as spin fragmentation \cite{Brooks-Bartlett_2014,Canals_2016}. Neutron-scattering measurements provide a direct experimental signature of spin fragmentation \emph{via} the coexistence of magnetic Bragg peaks and highly-structured magnetic diffuse scattering with pinch-point singularities \cite{Paddison_2016,Petit_2016,Lefrancois_2017}.  The divergence-full channel corresponds to an ``all-in/all-out" (AIAO) order of fragmented spins that reflects the long-range staggered arrangement of emergent charges. In the divergence-free channel, fragmented spins are disordered but correlated and the constraint that every triangle has zero emergent charge yields a Coulomb phase analogous to pyrochlore spin ices \cite{Brooks-Bartlett_2014}. Ultimately, similar to the pyrochlore spin ices, the dipolar interaction is expected to drive the system to a complete spin order at low temperature~\cite{Moller_2009,Chern_2011}, although such thermally equilibriated states are generally not realized experimentally in real systems.

Without additional quantum effects, the classical spin-fragmented (CSF) state described above can be viewed as a classical spin liquid coexisting with magnetic order. Its elementary excitations are thermally flipped spins that map onto pairs of magnetic monopoles, \textit{i.e.}, defects in the divergence-free channel. Typically, such thermally-activated excitations are exponentially suppressed at low temperature, as observed experimentally in the CSF phase of the kagome ice compound Dy${_3}$Mg${_2}$Sb${_3}$O${_{14}}$ \cite{Paddison_2016,Bai_unpublished}. The question arises as to  the effect of quantum fluctuations on degenerate classical spin configurations and whether a distinct phase of matter can be stabilized through quantum superposition, in close analogy to the quantum ice physics proposed for pyrochlore and square-lattice systems \cite{Hermele_2004,Savary_2012,Gingras_2014,Henry_2014,Savary_2017, Stern2019quantum}. For instance, quantum tunneling can connect different CSF configurations through the concurrent flipping of six spins, a process shown with a green hexagon in Fig.~\ref{fig1}. Conceptually, such dynamics may stabilize a putative ``quantum spin fragmented" (QSF) state that remains characterized by the coexistence of Bragg peaks and highly-structured diffuse scattering in magnetic neutron scattering experiments, but with dynamic magnetic correlations reflecting coherent collective excitations, akin to the emergent monopole and photon-like excitations in quantum spin ice \cite{Owen_2012,Gingras_2014}. It is unclear if such a partially disordered quantum state can theoretically prevail as an extended phase at finite temperature. Indeed, the previously-studied transverse-field Ising model on the kagome lattice maps onto a quantum dimer model for which cooperative quantum fluctuations exist only at a finely tuned Rokhsar-Kivelson point \cite{Moessner_2001, Misguich2002Quantum, Nikolic_2005}. Furthermore, the dipolar kagome ice model shows a robust tendency towards magnetic ordering at low temperatures~\cite{Chern_2011}, even when a small transverse-field is included~\cite{wang2020tuning}.

This rich theoretical landscape motivates comprehensive experimental investigation of dipolar kagome ice materials. 
In real systems, details of the magnetic Hamiltonian going beyond the necessarily-simplified aforementioned theoretical models will inevitably come into play. Some features, such as effective dimensionality of the magnetic interactions or sub-leading terms, may provide opportunities to increase quantum fluctuations with fine tuning. Other features, such as structural disorder or couplings to other degrees of freedom, usually suppress coherent quantum tunneling. Uncovering such effects and elucidating their role is a crucial and necessary step towards the experimental realization and engineering of genuinely quantum phases, such as the putative QSF state mentioned above.

In this work, we present comprehensive inelastic neutron scattering data which uncovers spin dynamics at the lowest measurable temperatures ($\approx 0.1$\,K) in the dipolar kagome Ising magnet Ho${_3}$Mg${_2}$Sb${_3}$O${_{14}}$ \cite{Dun_2017}. This material is one of a series of ``tripod kagome" materials derived from the pyrochlore structure by chemical substitution, yielding kagome planes of magnetic rare-earth ions separated by triangular planes of nonmagnetic Mg$^{2+}$ ions [Fig.~\ref{fig2}(a)]. Previous measurements of isostructural Dy${_3}$Mg${_2}$Sb${_3}$O${_{14}}$ revealed a CSF state at low temperature \cite{Paddison_2016}, in which no spin dynamics were observed in either neutron-scattering or ac-susceptibility data \cite{Dun_2016, Paddison_2016}. Our measurements on  Ho${_3}$Mg${_2}$Sb${_3}$O${_{14}}$ uncover a spin-fragmented state with instantaneous magnetic correlations closely resembling those of Dy${_3}$Mg${_2}$Sb${_3}$O${_{14}}$. Yet, we observe structured dynamic magnetic correlations at low temperature, indicating persistent spin dynamics in sharp contrast to the Dy$^{3+}$ compound. We show this stems from the low symmetry of the tripod-kagome structure and the non-Kramers nature of the Ho$^{3+}$ ion, the combination of which generates an effective local magnetic field transverse to the Ising magnetic dipole moments. The effective low-energy Hamiltonian for Ho${_3}$Mg${_2}$Sb${_3}$O${_{14}}$ thus maps onto an iconic model of \emph{quantum} magnetism---interacting Ising spins in a transverse magnetic field \cite{Wang_1968}. We use neutron-scattering experiments to determine this Hamiltonian, and employ a combination of exact diagonalization, field theoretic, and Monte Carlo methods to understand its spin correlations and excitations. Our calculations complement previous studies of the quantum kagome ice model \cite{Moessner_2000, Moessner_2001, Nikolic_2005, Carrasquilla_2015, Wu_2018} by including dipolar interactions to delineate the putative role played by transverse fields to stabilize partially-disordered quantum phases. In Ho${_3}$Mg${_2}$Sb${_3}$O${_{14}}$, however, the physics of Ho$^{3+}$ ions is profoundly affected by the strong nuclear hyperfine coupling, which eventually destroys coherent quantum effects between sites. As a result, Ho${_3}$Mg${_2}$Sb${_3}$O${_{14}}$ realizes a spin-fragmented state with predominantly single-ion quantum fluctuations. A key insight of our work is thus to determine the interplay of two effects---intrinsic transverse field and hyperfine coupling---on the quantum dynamics of a highly frustrated magnet.

Our paper is structured as follows. In Section~\ref{sec:Methods}, we summarize the experimental methods that we employ. In Section \ref{sec:EffectHam}, we present neutron-scattering measurements of the crystal-field excitations of Ho${_3}$Mg${_2}$Sb${_3}$O${_{14}}$ and of a structurally analogous but magnetically-dilute system (Ho$_{0.01}$La$_{0.99}$)${_3}$Mg${_2}$Sb${_3}$O${_{14}}$, and use these measurements to parameterize the spin Hamiltonian of Ho${_3}$Mg${_2}$Sb${_3}$O${_{14}}$. In Section~\ref{sec:HC}, we report heat-capacity measurements that identify a magnetic phase transition at $T^{\ast}=0.32$\,K accompanied by a large specific heat feature of coupled electronic-nuclear origin. In Section~\ref{sec:Inelstic}, we report low-temperature inelastic neutron-scattering measurements on polycrystalline samples of Ho${_3}$Mg${_2}$Sb${_3}$O${_{14}}$. They reveal that spin fragmentation occurs below $T^{\ast}$, and that low-energy spin excitations are structured in both momentum and energy space at the lowest measurable temperatures. In Section~\ref{sec:THEORY}, we use theoretical modeling to understand our data. 
Finally, we conclude in Section~\ref{sec:Conclusions} with a discussion of the general implications of our study.

\section{Methods}\label{sec:Methods}
Two different polycrystalline samples of Ho${_3}$Mg${_2}$Sb${_3}$O${_{14}}$ were prepared for this study, the first using a traditional solid state reaction method (referred to as s.s. sample), and the second using a sol-gel method (referred to as s.g. sample). For the s.s. sample, stoichiometric ratios of  Ho$_2$O$_3$ (99.9\%), MgO (99.99\%), and Sb$_2$O$_3$ (99.99\%) fine powder were carefully ground and reacted at a temperature of 1350$^\circ$C in air for 24 hours. This heating step was repeated until the amount of impurity phases as determined by X-ray diffraction was not reduced further. The synthesized s.s. sample contained a small amount of Ho$_3$SbO$_7$ impurity (2.29(18) wt\%), which orders antiferromagnetically at $T_{\text N}$ = 2.07\,K \cite{Fennell_2001}. This impurity can be removed by the sol-gel synthesis method. For this synthesis, stoichiometric amounts of Ho(NO$_3$)$_3$, Mg(NO$_3$)$_3$ (prepared by dissolving Ho$_2$O$_3$ and  MgO in hot diluted nitric acid solution), and antimony tartarate (prepared by dissolving Sb$_2$O$_3$ in hot tartaric acid solution) were first mixed in a beaker.  Citrate acid with a metal-to-citrate molar ratio of 1:2 was then added to the solution followed by a subsequent heating on a hot plate at 120$^\circ$C overnight to remove excessive water.  The obtained gel-like solution was slowly heated to 200$^\circ$C in a box furnace to decompose the nitrate, and was pyrolyzed at 600$^\circ$C for 10-12 h in air. The obtained powder was then ground, pressed into a pellet and re-heated at 1300$^\circ$C until a well-reacted crystalline powder was obtained. Heat-capacity measurements presented below show that the thermo-magnetic behavior of the two samples is almost identical, expect for a small peak around $2.1$\,K in the s.s. sample originating from the Ho$_3$SbO$_7$ impurity. A Ho-diluted (La-doped) sample of (Ho$_{0.01}$La$_{0.99}$)${_3}$Mg${_2}$Sb${_3}$O${_{14}}$ was also synthesized with the same sol-gel technique, with 99\% Ho$_2$O$_3$ replaced by La$_2$O$_3$ powder (99.9\%, baked at 900$^\circ$C overnight before use).

Low-temperature specific-heat measurements were performed on a Quantum Design Physical Properties Measurement System instrument using dilution refrigerator ($0.07 \leq T \leq 4$~K) and standard ($1.6 \leq T \leq 100$~K) probes. For the dilution refrigerator measurement, the powder samples were cold-sintered with Ag powder. The contribution of the Ag powder was measured separately and subtracted from the data. The lattice contribution to the heat capacity was estimated from measurements of the isostructural nonmagnetic compound La$_3$Mg$_2$Sb$_3$O$_{14}$.

Powder X-ray diffraction measurements were carried out with Cu K$\alpha$ radiation ($\lambda = 1.5418$\,\AA) in transmission mode. Powder neutron-diffraction measurements were carried out using the HB-2A high-resolution powder diffractometer \cite{Garlea_2010} at the High Flux Isotope
Reactor at Oak Ridge National Laboratory, with a neutron wavelength of 1.546\,\AA. Rietveld refinements of the crystal and magnetic structures were carried out using the FULLPROF suite of programs \cite{Rodriguez_1993}. Peak-shapes were modeled by Thompson-Cox-Hastings pseudo-Voigt functions, and backgrounds were fitted using Chebyshev polynomial functions. 

Inelastic neutron-scattering measurements on the s.s. sample of Ho${_3}$Mg${_2}$Sb${_3}$O${_{14}}$ were carried out using the Fine-Resolution Fermi Chopper Spectrometer (SEQUOIA) \cite{Granroth_2010} at the Spallation Neutron Source of Oak Ridge National Laboratory, and the Disk Chopper Spectrometer (DCS) \cite{Copley_2003} at the NIST Center for Neutron Research. For the SEQUOIA experiment, a $\sim$5\,g powder sample of Ho${_3}$Mg${_2}$Sb${_3}$O${_{14}}$ was loaded in an aluminum sample container and cooled to 4\,K with a closed-cycle refrigerator. Data were measured with incident neutron energies of $120$, $60$, and $8$\,meV. The same measurements were repeated for an empty aluminum sample holder and used for background subtraction. For the DCS measurements, the same sample was loaded in a copper can, filled with 10\,bar of helium gas at room temperature, and cooled to millikelvin temperatures using a dilution refrigerator. The measurements were carried out with an incident neutron energy of $3.27$\,meV~at temperatures between $0.12$ and $40$\,K. Measurements of an empty copper sample holder were also made and used for background subtractions. Due to the large specific heat and related relaxation processes below $1$\,K, a thermal stabilization time of $6$\,h was used; no change in the data was observed after this waiting time. Data reduction was performed using the DAVE program \cite{Azuah_2009}. For modeling and fitting purposes, data were corrected for background scattering using empty-container measurements and/or high-temperature measurements, as specified in the text. These data were also corrected for neutron absorption \cite{howard1987determination}, placed on an absolute intensity scale by scaling to the nuclear Bragg profile, and the magnetic scattering from the Ho$_{3}$SbO$_{7}$ impurity below its $T_{\text N}$ of $2.07$\,K was subtracted as described in Ref.~\onlinecite{Paddison_2016}. Additional inelastic neutron-scattering measurements on Ho${_3}$Mg${_2}$Sb${_3}$O${_{14}}$ (s.g. sample) and (Ho$_{0.01}$La$_{0.99}$)${_3}$Mg${_2}$Sb${_3}$O${_{14}}$ were carried out using the OSIRIS backscattering spectrometer at the ISIS neutron source with a final neutron energy of $1.84$\,meV. For (Ho$_{0.01}$La$_{0.99}$)${_3}$Mg${_2}$Sb${_3}$O${_{14}}$, a $13.6$\,g powder sample was loaded into an aluminum can and was cooled with an Orange cryostat to the base temperature of $1.6$\,K. For Ho${_3}$Mg${_2}$Sb${_3}$O${_{14}}$, $4$\,g of s.g. powder was wrapped in a thin copper foil and placed into a copper sample can with a capillary that allowed helium filling at low temperature. The system was cooled using a dilution refrigerator, and with a maximum of 3\,bar filled helium, the lowest sample temperature accessed in this experiment was estimated to be 400$\pm50$\,mK by comparison with the DCS data.  

For convenience, in the following sections, we use a unit system with $k_\mathrm{B}=1$ and $\hbar=1$, so that all energies are given in units of K.

\section{Effective Hamiltonian} \label{sec:EffectHam}

\subsection{Crystal structure and interactions}

\begin{figure*}[tbp] 
	\begin{center}
		\includegraphics[width= 7 in]{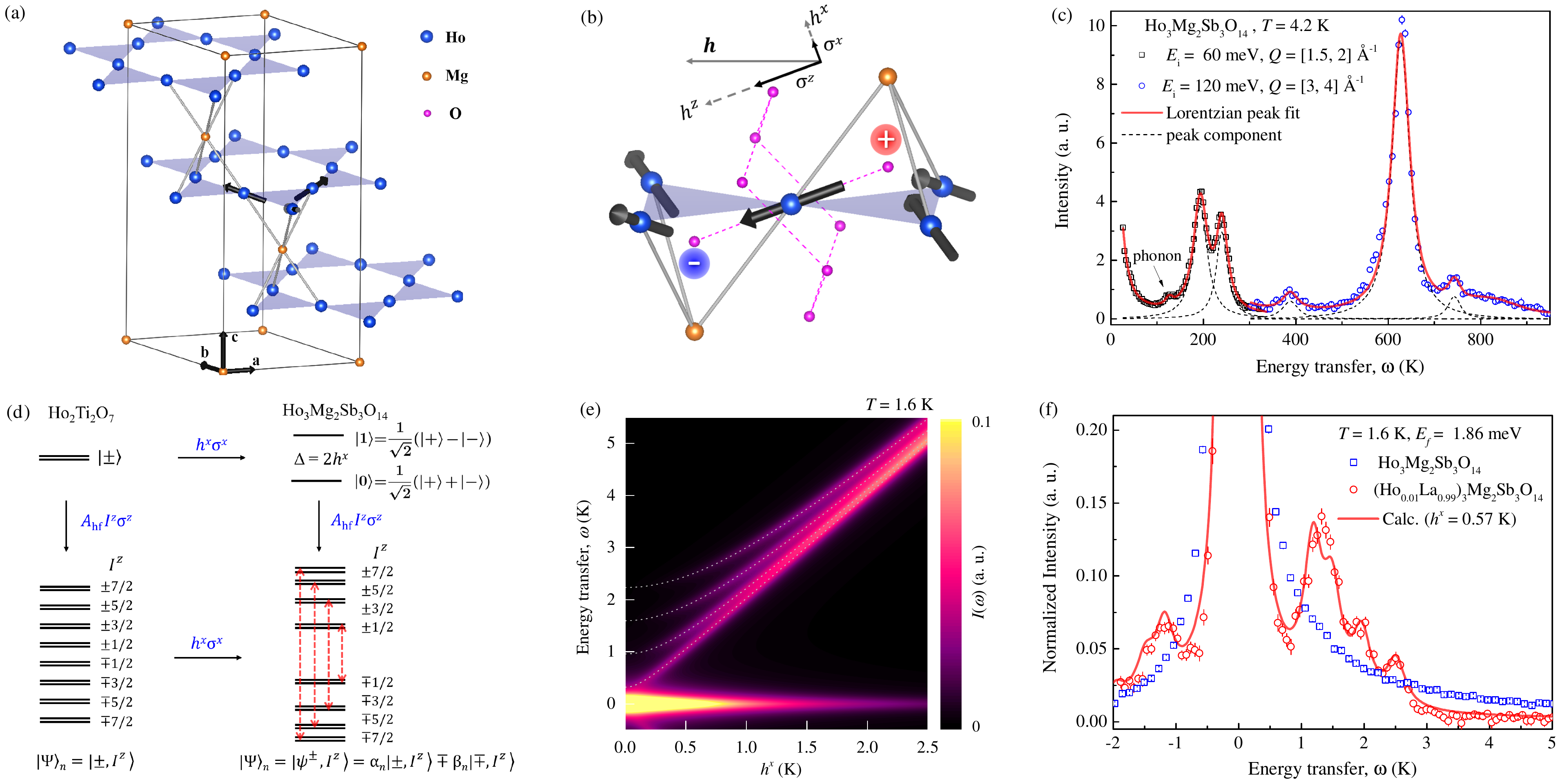}
	\end{center}
	\caption{\label{fig2}{\bf } 
(a) Simplified partial crystal structure of Ho${_3}$Mg${_2}$Sb${_3}$O${_{14}}$, showing alternating Ho$^{3+}$ kagome layers (large blue spheres) and Mg$^{2+}$ triangular layers (small orange spheres). (b)  Local environment of  Ho$^{3+}$ ions, showing eight oxygen ions (small pink spheres), two Mg$^{2+}$ ions, and four nearest-neighbor Ho$^{3+}$ ions around a central Ho$^{3+}$ ion. The tripod-like arrangement of the local Ising axes is enforced by the oxygen in the center of each MgHo$_3$, which yields a canting angle of 22.3$^\circ$ with respect to the kagome plane. Spins are labeled by black arrows while $Q_j=+1$ and $Q_j=-1$ magnetic charges are illustrated by red and blue spheres, respectively. The coordinate system depicts the local Ising, transverse field and mean-field directions. (c) Crystal-field excitations measured by inelastic neutron-scattering experiments on the SEQUOIA spectrometer. Open black circles and blue squares indicate intensities measured with incident neutron energies of $60$ and $120$\,meV, respectively. Five crystal-field levels can be directly resolved at energies of 190(2), 238(2), 388(9), 627(5), and 743(9)\,K, whose peak positions are extracted from Lorentzian fits to the data (black dashed lines). 
The overall fit, including a modeled phonon background, is shown as a red line. Detailed analysis of the crystal field excitations can be found in a separate study \cite{Dun_2020}.   (d) Comparison of the single-ion response expected for a Ho$^{3+}$ ion in a pyrochlore \emph{vs.} a tripod kagome geometry. Crystal-field singlets $\ket{0}$ and $\ket{1}$ in Ho${_3}$Mg${_2}$Sb${_3}$O${_{14}}$ can be effectively described as a transverse field $h^x$ acting on the non-Kramers doublet $\ket{\pm}$ of Ho$_2$Ti$_2$O$_7$. While hyperfine interactions split the electronic-nuclear manifold into eight levels in both systems, the transverse field $h^x$   transforms the classical single-ion picture  of Ho$_2$Ti$_2$O$_7$ into a quantum picture in Ho${_3}$Mg${_2}$Sb${_3}$O${_{14}}$, where each eigenstate is a quantum superposition of $\ket{+, I^z}$ and $\ket{-, I^z}$. Red dashed arrows mark the allowed transitions between eigenstates of the same $I^z$. (e) Calculated scattering intensity of single-ion excitations as a function of $h^x$ at a temperature of 1.6\,K. Dashed lines present the transition energies whose intensities are calculated and convoluted with a Lorentzian function with FWHM = 0.35\,K. (f) Low-energy inelastic neutron scattering response measured on the OSIRIS spectrometer for pure Ho${_3}$Mg${_2}$Sb${_3}$O${_{14}}$ (s.g. sample) and the dilute Ho-tripod magnet (Ho$_{0.01}$La$_{0.99}$)${_3}$Mg${_2}$Sb${_3}$O${_{14}}$ at a temperature of 1.6\,K. The two datasets are normalized to the same total spectrum weight. Solid lines represent the single-ion spectrum calculated using $h^x = 0.57$\,K, which yields the best agreement with the experimental data.   }
\end{figure*} 

The crystal structure of Ho${_3}$Mg${_2}$Sb${_3}$O${_{14}}$ (space group $R\bar{3}m$) is shown in Fig.~\ref{fig2}(a), and contains kagome planes of magnetic Ho$^{3+}$ ions separated by triangular layers of nonmagnetic Mg$^{2+}$ \cite{Dun_2017}. The Ho$^{3+}$ site has $C_{2h}$ point symmetry and its local environment contains eight oxygen atoms \cite{Dun_2016, Dun_2017}. The orientations of Ho$^{3+}$ magnetic dipole moments are constrained by crystal electric field effects to point along the line connecting Ho$^{3+}$ to its two closest oxygen neighbors, which are situated near the centroids of the MgHo$_3$ tetrahedra [Fig.~\ref{fig2}(b)]. Rietveld co-refinements to X-ray and neutron powder-diffraction data for Ho${_3}$Mg${_2}$Sb${_3}$O${_{14}}$ (s.s. sample) confirm this crystal structure, and reveal a small amount of Ho$^{3+}$/Mg$^{2+}$ site mixing such that $3.2(2)$\% of Ho$^{3+}$ atomic positions are occupied in a disordered way by Mg$^{2+}$ (see Appendix~A). Hence, the extent of chemical disorder in Ho${_3}$Mg${_2}$Sb${_3}$O${_{14}}$ is less than in its Dy$^{3+}$ analog, where the corresponding value is 6(2)\% \cite{Paddison_2016}. 

We anticipate that the spin Hamiltonian for Ho${_3}$Mg${_2}$Sb${_3}$O${_{14}}$ may be written as a sum of three terms,
\begin{equation}
    \mathcal{H}=\mathcal{H}_\mathrm{cf}+\mathcal{H}_\mathrm{hf} +\mathcal{H}_\mathrm{int}, \label{eq:h1}
\end{equation}
where $\mathcal{H}_\mathrm{cf}$, $\mathcal{H}_\mathrm{hf}$, and $\mathcal{H}_\mathrm{int}$ denote respectively the crystal-field, nuclear hyperfine, and pairwise interaction Hamiltonian. We now consider the origin, form, and magnitude of each term, and show that they are all relevant in Ho${_3}$Mg${_2}$Sb${_3}$O${_{14}}$ at low temperature.

\subsection{Crystal-field Hamiltonian}\label{sub:cef}

We use inelastic neutron-scattering measurements and point-charge calculations to determine the parameters of $\mathcal{H}_\mathrm{cf}$. The high-energy spectrum observed in  Ho${_3}$Mg${_2}$Sb${_3}$O${_{14}}$ (s.s. sample) comprises five crystal-field excitations, with energies and relative intensities that resemble those of pyrochlore spin ice Ho$_2$Ti$_2$O$_7$ \cite{Rosenkranz_2000,Ruminy_2016} except for an overall downwards renormalization in energy [Fig.~\ref{fig2}(c)]. This overall resemblance is expected given the similar local environments for Ho$^{3+}$ ions in these two systems. However, a crucial difference stems from the reduced $C_{2h}$  symmetry of the Ho$^{3+}$ site in Ho${_3}$Mg${_2}$Sb${_3}$O${_{14}}$ compared to the $D_{3d}$ symmetry in Ho$_2$Ti$_2$O$_7$. Whereas the crystal-field ground-state in Ho$_2$Ti$_2$O$_7$ is a non-Kramers doublet, in Ho${_3}$Mg${_2}$Sb${_3}$O${_{14}}$ all crystal-field levels are necessarily singlets \cite{Dun_2017}. 
However, as a probable consequence of spin-spin interactions (see Section~\ref{sec:Inelstic}), the excitation associated with the splitting of the ground-state doublet is strongly overdamped [Fig.~\ref{fig2}(f)]. This precludes a direct neutron-scattering measurement of this energy splitting in Ho${_3}$Mg${_2}$Sb${_3}$O${_{14}}$. We addressed this problem using two complementary approaches. First, we performed point-charge calculations using an effective charge model that matches the high-energy crystal-field excitation spectrum (see Appendix B) \cite{Dun_2020}. Second, we validated the low-energy predictions of this model using high-resolution neutron scattering measurements on a magnetically-dilute Ho tripod kagome compound, (Ho$_{0.01}$La$_{0.99}$)${_3}$Mg${_2}$Sb${_3}$O${_{14}}$, such that interaction effects between sites are negligible. We discuss these results in turn below.

Our point-charge model \cite{Dun_2020} predicts that the ground-state doublet is split into two singlets separated by an energy gap $\Delta\approx1.74$\,K (see Appendix~B). The two singlets are well approximated by symmetric and anti-symmetric superpositions of pure free-ion states,
\begin{eqnarray}
\ket{0} \approx \frac{1}{\sqrt{2}} (\ket{+}+\ket{-}), \nonumber\\
\ket{1} \approx \frac{1}{\sqrt{2}} (\ket{+}-\ket{-}), \label{eq:singlets}
\end{eqnarray} 
where $\ket{\pm}\!\equiv\ket{J\!=\!8,J^z\!=\!\pm8}$ represents the ground-state doublet of Ho$^{3+}$ in Ho$_2$Ti$_2$O$_7$ if we ignore the small contributions from other $J^z$ components \cite{Rosenkranz_2000,Rau_2015}. At low temperatures, only $\ket{0}$ and $\ket{1}$ are thermally populated because of their $190$\,K separation from higher-energy crystal-field levels [Fig.~\ref{fig2}(c)], which is consistent with higher-temperature inelastic neutron data \cite{Dun_2020} and specific-heat measurements [Section~\ref{sec:HC}]. The form of $\ket{0}$ and $\ket{1}$ allows for a nonzero angular momentum matrix element  $\langle0|\hat{J}^{z}|1\rangle 
\approx$ 8 while  $\langle0|\hat{J}^{\alpha}|1\rangle$ with ${\alpha = x, y}$ are vanishingly small (more accurate values are given in  Appendix~B).
Similar to the procedure for pyrochlore spin ices \cite{Rau_2015}, we construct Pauli matrices using the two thermally-accessible crystal-field states,
\begin{eqnarray}
\sigma^{\pm} = 2\ket{\pm}\bra{\mp}, \;\; \sigma^{z} = \ket{+}\bra{+}-\ket{-}\bra{-}. \label{eq:pseudospin}
\end{eqnarray} 
In this framework, the total magnetic dipole moment operator is related to $\sigma^z$ as
\begin{equation} 
\boldsymbol{\mu}= -g_J\mu_\mathrm{B} \sigma^{z} \langle0|\hat{J}^{z}|1\rangle  \hat{\mathbf{z}}, \label{eq:Moment}
\end{equation}
where $g_J=\frac{5}{4}$ is the Ho$^{3+}$ Land\'{e} factor, and $\hat{\mathbf{z}}$ is a local Ising axis shown in Fig.~\ref{fig2}(b). The nonzero matrix element $\langle0|\hat{J}^{z}|1\rangle $ is therefore expected to generate a total magnetic moment of magnitude  $9.74\,\mu_\mathrm{B}$, which is conserved. In contrast, static moments only appear when $\langle\sigma^{z}\rangle$ becomes non-zero, \emph{e.g.} under an external magnetic field. Meanwhile, $\sigma^x$ and $\sigma^y$ transform  as higher order multipoles which are not directly observable in our neutron scattering measurements. These predictions are supported by our isothermal magnetization measurements between $1.8$\,K and $40$\,K, which are consistent with a model of paramagnetic Ising spins  (see Appendix~C). 

It is established \cite{Wang_1968,Savary_2017} that an energy splitting between two crystal-field singlets can be exactly mapped into a transverse magnetic field acting on a corresponding doublet. This mapping can be understood by recognizing that $\ket{0}$ and $\ket{1}$ are the eigenstates of the  $\sigma^x=\small{\begin{pmatrix} 0 & 1 \\ 1 & 0 \end{pmatrix}}$ Pauli matrix. Therefore, in the Pauli matrix picture, our crystal field Hamiltonian can be recast as
\begin{equation}
\mathcal{H}_{\mathrm{cf}}=h^{x}\sigma^x,\label{eq:CF}
\end{equation}
where $h^x=\Delta/2$ is the intrinsic transverse field [Fig.~\ref{fig2}(d)]. We will use this pseudo-spin representation in the rest of this paper.

\subsection{Nuclear hyperfine Hamiltonian}
Hyperfine interactions couple non-zero nuclear spins to the local magnetic field from surrounding electrons. With a nuclear spin quantum
number $I=7/2$, $^{165}$Ho is the only stable isotope of holmium and its hyperfine energy-scale is the strongest among the rare-earth elements. For a non-Kramers electronic system in the pseudo-spin approximation, the hyperfine Hamiltonian takes the simple form \cite{abragam2012electron}
\begin{equation}
\mathcal{H}_{\mathrm{hf}}=A_\mathrm{hf}I^z\sigma^z \label{eq:hyperfine}
\end{equation}
where $I^z = -I, ..., I$ labels the $z$-component of the nuclear spin operator, and $A_\mathrm{hf}= 0.319$\,K is the hyperfine coupling constant for Ho \cite{Krusius_1969,kondo1961internal}. We neglect the electric quadrupole coupling constant because its energy scale ($P=0.004$\,K) is very small.

In Ho$_2$Ti$_2$O$_7$, the hyperfine coupling splits the combined electronic and nuclear system into eight uniformly-spaced levels; each level is doubly degenerate because of the Kramers degeneracy of the combined electronic and nuclear spin system [Fig.~\ref{fig2}(d)]. At the single-ion level, the system remains classical because both  $I^z$ and $\sigma^z$ are good quantum numbers, and eigenstates can be labeled as $\ket{\pm,I^z}$. By contrast, for Ho${_3}$Mg${_2}$Sb${_3}$O${_{14}}$, the single-ion Hamiltonian contains both an intrinsic transverse field and hyperfine interactions,
\begin{equation}
 \mathcal{H}_{\mathrm{SI}}=\mathcal{H}_{\mathrm{cf}}+\mathcal{H}_{\mathrm{hf}}.\label{eq:single-ion}
\end{equation} 
This generates a dynamic electronic moment since neither $\sigma^z$ nor $\sigma^x$ are good quantum numbers. Diagonalizing the 16-dimensional 
$\mathcal{H}_{\mathrm{SI}}$ yields an unevenly spaced spectrum for which each eigenstate is a quantum superposition of $\ket{+,I^z}$ and $\ket{-,I^z}$ whose mixing depends on the values of $I^z$ and  $h^x$, and the lowest energy states always have $I^z=\pm7/2$. To illustrate this effect, Fig.~\ref{fig2}(e) shows the low-temperature magnetic scattering intensity of electronic spins as a function of $h^x$.  For vanishing $h^x$, the magnetic response is purely elastic because the dipolar matrix elements connecting different eigenstates are zero. With increasing $h^x$, the scattering acquires an inelastic component that comprises four excitations, arising from transitions between eigenstates with the same $I^z$ [Fig.~\ref{fig2}(d)]. Finally, in the limit of $h^x\gg A_\mathrm{hf}I$, a single crystal-field excitation at $\Delta = 2 h^x$ is obtained.
	
To test our model single-ion Hamiltonian against experiment, we employ high-resolution neutron scattering measurements of the magnetically-dilute Ho tripod kagome compound (Ho$_{0.01}$La$_{0.99}$)${_3}$Mg${_2}$Sb${_3}$O${_{14}}$. We assume that the Ho$^{3+}$ ions are randomly distributed on the kagome lattice and pairwise interactions between them are negligible. Neutron-scattering data measured at 1.6\,K are shown in Fig.~\ref{fig2}(f); they display a broad and intense peak at $\omega=1.3$\,K, along with two narrower and weaker side peaks at $\omega=1.9$\,K and $2.5$\,K. The experimental data are in excellent quantitative agreement with exact diagonalization calculations of $H_\mathrm{SI}$, taking $h^x=0.57$\,K [Fig.~\ref{fig2}(e)]. This result yields strong evidence that $H_\mathrm{SI}$ describes well the single-ion properties of Ho-based tripod kagome magnets. Importantly, due to the larger ionic size of La$^{3+}$ compared with Ho$^{3+}$, the lattice parameters of (Ho$_{0.01}$La$_{0.99}$)${_3}$Mg${_2}$Sb${_3}$O${_{14}}$ are approximately 3\% larger than those of Ho${_3}$Mg${_2}$Sb${_3}$O${_{14}}$. Using a power law scaling of $h^x$ as a function of lattice parameters, we extrapolate to $h^x = 0.85$\,K in Ho${_3}$Mg${_2}$Sb${_3}$O${_{14}}$ \cite{Dun_2020}, which is in good agreement with the point charge estimate of $h^x = 0.77$\,K. We take $h^x = 0.85$\,K for Ho${_3}$Mg${_2}$Sb${_3}$O${_{14}}$ throughout the rest of this paper.

\begin{figure} 
	\begin{center}
		\includegraphics[width= 3 in]{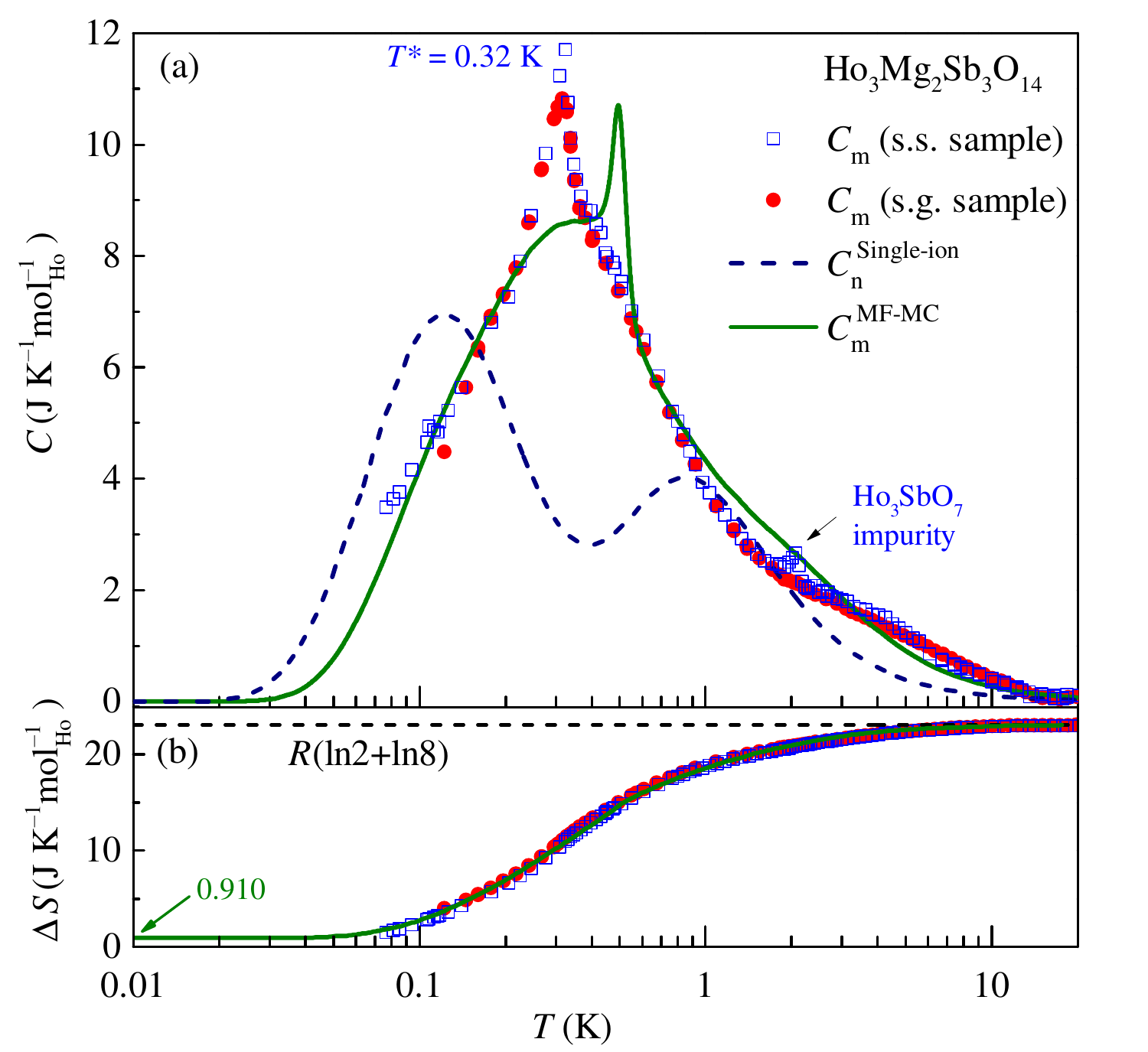}
	\end{center}
	\caption{\label{fig3}  (a) Magnetic contribution to the specific heat in Ho${_3}$Mg${_2}$Sb${_3}$O${_{14}}$ ($C_{\text{m}}$, shown as blue squares for the s.s. sample and red dots for the s.g. sample). The specific heat contains coupled electronic and nuclear spin contributions.  The navy dashed line shows the single-ion nuclear specific heat calculated assuming an isolated Ho$^{3+}$ ion with $h_x = 0.85$\,K and $A_\mathrm{hf} = 0.319$\,K.  Green solid lines show a mean-field Monte Carlo calculation including spin-spin interactions ($J_{\text{nn}} = -0.64$\,K, $D = 1.29$\,K). (b) Magnetic entropy change for $C_{\text{m}}$ from 20\,K to the lowest measurement temperature of $76$\,mK (colors and symbols as above). Our measurements on the s.s. sample render a residual entropy of 1.5(5) $\mathrm{J\,K}^{-1}\mathrm{mol_{Ho}^{-1}}$ at 87\, mK, and our mean-field Monte Carlo simulations predict a zero-point entropy of 0.910 $\mathrm{J\,K}^{-1}\mathrm{mol_{Ho}^{-1}}$.} 
\end{figure}

\subsection{Transverse Ising model}

We now consider the effect of pairwise interactions $J_{ij}$  between Ho$^{3+}$ ions. 
By analogy with spin-ice pyrochlores \cite{Bramwell_2001} and isostructural Dy${_3}$Mg${_2}$Sb${_3}$O${_{14}}$ \cite{Paddison_2016}, we expect that $J_{ij}$ contains a combination of nearest-neighbor exchange interactions $J_\mathrm{nn}$ and long-range magnetic dipolar interactions of overall scale $D$.  In principle, interactions between transverse spin components are also possible, but they are expected to be several orders of magnitude smaller  \cite{Rau_2015}, so we do not consider them further. The pairwise interaction Hamiltonian is therefore
\begin{equation}\label{eq:TIM}
\mathcal{H}_\mathrm{int}  =   \frac{1}{2}\sum_{i,j}J_{ij}\sigma_i^z\sigma_j^z,
\end{equation}
with
\begin{equation}
J_{ij} =J_{\mathrm{nn}}\delta_{r_{ij},r_\mathrm{nn}}+Dr_{\mathrm{nn}}^{3}\frac{\hat{\mathbf{z}}_{i}\cdot\hat{\mathbf{z}}_{j}-3(\hat{\mathbf{z}}_{i}\cdot\hat{\mathbf{r}}_{ij})(\hat{\mathbf{z}}_{j}\cdot\hat{\mathbf{r}}_{ij})}{r_{ij}^{3}},  \label{eq:Jij}
\end{equation}
where $\delta_{r_{ij},r_\mathrm{nn}}$ is the Kronecker delta function, $r_\mathrm{nn}$ is the distance between nearest-neighbor Ho$^{3+}$ ions, $r_{ij}$ is the distance between ions at positions $\mathbf{r}_i$ and $\mathbf{r}_j$, and $\hat{\mathbf{r}}_{ij}=(\mathbf{r}_{i}-\mathbf{r}_{j})/r_{ij}$. The value of $D =  \mu_{0}\mu^{2}/({4\pi}{k_{\mathrm{B}}r_{\mathrm{nn}}^{3}})=1.29$\,K is  fixed by the crystal structure, and we will obtain an experimental estimate of $J_\mathrm{nn}\approx-0.64$\,K in Section~\ref{sec:Inelstic}.
With  $\hat{\mathbf{z}}_{i}\cdot\hat{\mathbf{z}}_{j}=-0.28$, and $-3(\hat{\mathbf{z}}_{i}\cdot\hat{\mathbf{r}}_{nn})(\hat{\mathbf{z}}_{j}\cdot\hat{\mathbf{r}}_{nn})=1.93$, the sum of exchange and dipolar couplings at the nearest neighbor level 
is approximately $1.69$\,K, and hence antiferromagnetic in the pseudo-spin language. 
Consequently, magnetic interactions between sites are frustrated, and furthermore, comparable in magnitude to $h^x$.

Using the results of the previous subsections, we rewrite the full spin Hamiltonian, Eq.~\eqref{eq:h1}, as

\begin{equation}\label{eq:EffectiveH}
\mathcal{H} =  A_\mathrm{hf}\sum_iI_i^z\sigma_i^z + h^{x}\sum_i \sigma_i^x  +  \frac{1}{2}\sum_{i,j}J_{ij}\sigma_i^z\sigma_i^z.
\end{equation}
Eq.~\eqref{eq:EffectiveH} is equivalent in form to 
an Ising model in a transverse field (TIM), with an additional on-site longitudinal field due to the hyperfine coupling. The TIM has been used to model diverse physical phenomena, including ferroelectricity \cite{Brout_1966,Stinchcombe_1973}, superconductivity \cite{Anderson_1958}, quantum information \cite{Suzuki_2012,Dutta_2015}, and quantum phase transitions \cite{Ronnow_2005, Coldea_2010}. Typically, the pairwise interactions that drive magnetic ordering compete with the transverse field that drives quantum tunneling. In the absence of geometrical frustration, a phase transition only occurs to a magnetically-ordered state if $J_{ij}$ dominates over $h^x$. The interplay of frustration and transverse field may generate exotic quantum phases \cite{Moessner_2000, Moessner_2001, Nikolic_2005,Savary_2017}. On the kagome lattice, the TIM with nearest-neighbor antiferromagnetic interactions is predicted to have a quantum-disordered ground state for small $h^x$, smoothly connected to a quantum paramagnetic state at large $h^x$ \cite{Moessner_2000, Moessner_2001, Nikolic_2005}. 
On the pyrochlore lattice, an external field cannot be applied transverse to all spins simultaneously because the different local Ising axes are not coplanar, and a homogeneous transverse field that emerges at single-ion level is absent in chemically-ordered pyrochlores. However, transverse fields generated by spin-spin interactions are related to monopole hoping in spin ice \cite{Tomasello_2019}, and transverse fields generated by chemical disorder have been identified as a possible route to pyrochlore QSL states \cite{Savary_2017}, and used to explain the spin dynamics of Pr$_2$Zr$_2$O$_7$  \cite{Wen_2017, Sibille_2018} and Tb$_2$Ti$_2$O$_7$ \cite{Bonville_2011,Petit_2012}. Nevertheless, a potential challenge to modeling such materials is that chemical disorder generates a broad distribution of transverse fields in the sample \cite{Benton_2017}. Hence, a key feature of Ho${_3}$Mg${_2}$Sb${_3}$O${_{14}}$ is that its transverse field is intrinsic to the chemically-ordered structure, and is homogeneous to a first approximation.

\section{Specific-heat measurements}\label{sec:HC}
We use heat capacity measurements to understand thermodynamic properties of Ho${_3}$Mg${_2}$Sb${_3}$O${_{14}}$ and identify possible phase transitions. A sharp peak in the magnetic specific heat ($C_\mathrm{m}$) is observed at $T^\ast = 0.32$\,K for both s.s. and s.g. samples, indicating a symmetry-breaking magnetic phase transition [Fig.~\ref{fig3}(a)]. The value of $T^\ast $ is consistent with the broad peak previously observed around 0.4\,K using the ac susceptibility technique \cite{Dun_2017}. Whereas the ac susceptibility peak is frequency dependent \cite{Dun_2017}, the sharpness of the $C_\mathrm{m}$ peak is inconsistent with a conventional spin freezing scenario. The value of $T^\ast$ is also close to the temperature at which the isostructural compound Dy${_3}$Mg${_2}$Sb${_3}$O${_{14}}$ undergoes a phase transition from a kagome spin-ice state to a CSF state ($\sim$$0.3$\,K in Ref.~\onlinecite{Paddison_2016} and $\sim$$0.37$\,K in Ref.~\onlinecite{Dun_2016}). As we will show in Section~\ref{sec:Inelstic}, $T^\ast$ corresponds to the onset of a spin-fragmented state in Ho${_3}$Mg${_2}$Sb${_3}$O${_{14}}$,  characterized by a reduced ordered moment compared to a CSF state.

Below $1$\,K, a broad specific-heat feature is observed in addition to the sharp peak, consistent with a nuclear spin contribution. By integrating $C_\mathrm{m}/T$ from 20\,K to the lowest measurement temperature of 76\,mK, the recovered magnetic entropy reaches 21.5(5)\,$\mathrm{J\,K}^{-1}\mathrm{mol_{Ho}^{-1}}$, which is 6.5\% smaller than the expectation of $R(\ln2+\ln8) = 23.05\, \mathrm{J\,K}^{-1}\mathrm{mol_{Ho}^{-1}}$ considering both the electronic and nuclear spin degrees of freedom [Fig.~\ref{fig3}(b)] and assuming the absence of low-lying excited crystal-field states.
Notwithstanding the systematic uncertainty arising from the large nuclear specific heat, the experimental residual entropy of $1.5(5)$\,$\mathrm{J\,K}^{-1}\mathrm{mol_{Ho}^{-1}}$ at 87\,mK is comparable to the expected residual entropy of $0.92$\,$\mathrm{J\,K}^{-1}\mathrm{mol_{Ho}^{-1}}$ associated with the CSF degeneracy of the electronic spins \cite{Chern_2011, Paddison_2016}, consistent with a spin fragmentation picture in Ho${_3}$Mg${_2}$Sb${_3}$O${_{14}}$. 

In Ho-based systems with a doublet single-ion ground state, such as Ho metal \cite{Krusius_1969}, Ho$_2$Ti$_2$O$_7$ \cite{Bramwell_2001PRL}, and LiHoF$_4$ \cite{MENNENGA_1984}, the nuclear specific heat consists of a broad peak (Schottky anomaly) below the ordering temperature of the electronic spins. 
This implies that the dynamics of the electronic and nuclear subsystems separate in such systems, with electronic spins already in their $\sigma^z$ eigenstate when nuclear spins start to follow them at low temperature. This paradigm is not applicable to Ho${_3}$Mg${_2}$Sb${_3}$O${_{14}}$ because $h^x$ mixes $\ket{+, I^z}$ and  $\ket{-, I^z}$ at each site [Fig.~\ref{fig2}(d)]. Accordingly, the single-ion Hamiltonian [Eq.~\eqref{eq:single-ion}] predicts two peaks for the nuclear specific heat; however, this model strongly disagrees with our experimental data  [Fig.~\ref{fig3}(a)].
The observed $C_\mathrm{m}$ is also very different from that of unfrustrated TIM magnets such as HoF$_3$, in which hyperfine interactions act as an effective mean field that precipitates long-range ordering of electronic spins, and the nuclear Schottky anomaly is only manifest below $T^\ast$ \cite{Ramirez_1994}. By contrast, in Ho${_3}$Mg${_2}$Sb${_3}$O${_{14}}$, more than half of the total magnetic entropy has already been recovered above $T^\ast$, suggesting the development of short-range spin correlations with a coupled electronic-nuclear character.
Our heat capacity measurements thus provide the first experimental hint of many-body physics in Ho${_3}$Mg${_2}$Sb${_3}$O${_{14}}$, demonstrating that the hyperfine term, transverse field, and spin-spin interactions in Eq.~\eqref{eq:EffectiveH} must be treated on an equal footing.

\section{Inelastic neutron-scattering measurements}\label{sec:Inelstic}
\begin{figure*}[tbp]
	\begin{center}
		\includegraphics[width= 7 in]{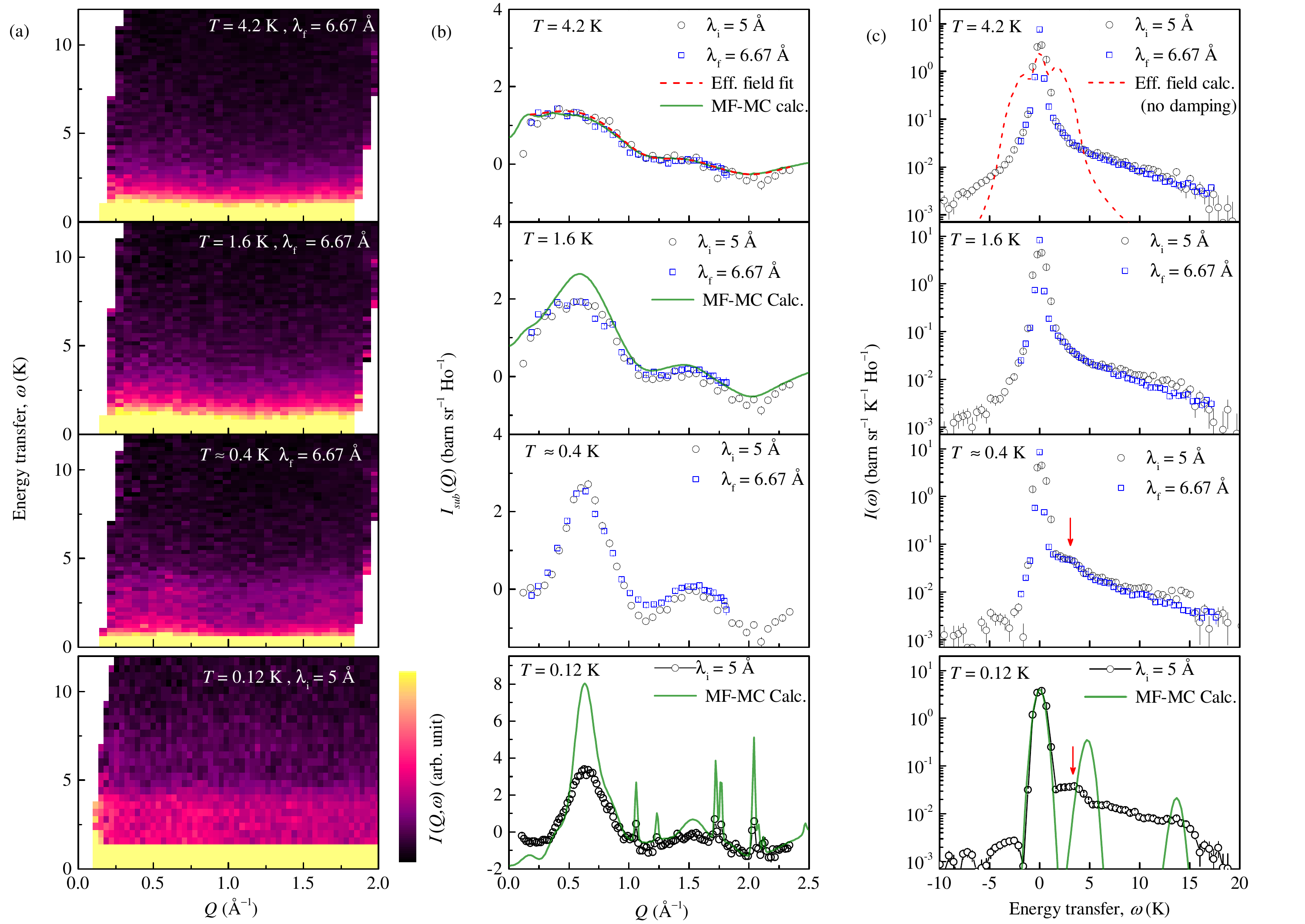}
	\end{center}
	\caption{\label{fig4}  Low-energy neutron-scattering response of Ho${_3}$Mg${_2}$Sb${_3}$O${_{14}}$ at $T=4.2$\,K, $1.6$\,K, $0.4$\,K, and $0.12$\,K (top to bottom panels). Data are from two distinct measurements on different spectrometers and samples. Black circles correspond to measurements on the s.s. sample using the DCS spectrometer with an incident neutron wavelength of $5$\,\AA; blue squares correspond to measurements on the s.g. sample using the OSIRIS spectrometer with a final neutron wavelength of $6.67$\,\AA. (a) Dependence of the inelastic magnetic scattering intensity $I(Q,\omega)$ on wavevector transfer $Q$ and energy transfer $\omega$. 
	(b) Energy-integrated ($-20 \leq \omega \leq  20$\,K) magnetic neutron-scattering intensity $I_\mathrm{sub}(Q)$ at four temperatures, showing experimental data (black circles), fits using the paramagnetic effective field theory (dashed red lines), and calculations using the mean-field theory (solid olive lines). The correlated magnetic scattering has been isolated by subtracting high-temperature data or simulations taken at $T = 30$\,K, which leads to negative values for the subtracted intensity.
	(c) Energy dependence of magnetic neutron-scattering intensity $I(\omega)$ integrated over wavevector transfers $0.4 \leq Q \leq1.6 $\,\AA$^{-1}$ at four different temperatures. An empty container subtraction was used to remove background.}
\end{figure*}

We use inelastic neutron-scattering measurements to probe the spin correlations and low-energy spin dynamics of Ho${_3}$Mg${_2}$Sb${_3}$O$_{14}$. Neutron-scattering data as a function of momentum ($Q$) and energy transfer ($\omega$) are shown in Fig.~\ref{fig4}(a) over the temperature range from 4.2\,K to 0.12\,K. The $Q$-dependence of the magnetic scattering shown in Fig.~\ref{fig4}(b) was obtained by integrating over $-20 \leq \omega \leq  20$\,K  and subtracting the paramagnetic data at $T = 30$\,K. Such energy-integrated data measure the Fourier transform of the instantaneous spin-pair correlation function. The energy dependence shown in Fig.~\ref{fig4}(c) was obtained by integrating the inelastic scattering over $0.4 \leq Q \leq 1.6$\,\AA$^{-1}$ and correcting for background scattering using empty-container measurements.

We first discuss the paramagnetic regime above $T^\ast$. For sample temperatures between $4.2$\,K and $0.4$\,K, the energy-integrated magnetic diffuse scattering displays a clear $Q$-dependence, with a broad peak centered at approximately 0.65\,\Ang~that develops on cooling [Fig.~\ref{fig4}(b)]. This feature closely resembles observations for Dy${_3}$Mg${_2}$Sb${_3}$O$_{14}$, where it was interpreted in terms of the development of kagome-ice correlations with a ``one in, two out" or ``two in, one out" ice rule on each triangle \cite{Paddison_2016}. The energy-resolved response shows two main magnetic features: an intense and resolution-limited quasielastic peak; and a broad inelastic tail extending to $\omega\approx15$\,K that decays slowly with increasing energy transfer [Fig.~\ref{fig4}(c)]. This energy-resolved response differs dramatically from the case of Dy${_3}$Mg${_2}$Sb${_3}$O$_{14}$, in which only elastic ($\omega=0$) neutron scattering was observed at comparable temperatures, indicating time-independent spin correlations \cite{Paddison_2016}. Moreover, the energy-resolved response of Ho${_3}$Mg${_2}$Sb${_3}$O$_{14}$ is very different from the single-ion model discussed in Section~\ref{sec:EffectHam}, in which the  majority of scattering is expected to be inelastic and concentrated around energy transfer $\Delta=2h^{x}\approx1.7$\,K [Fig.~\ref{fig2}(e) and (f)]. The most likely reason for this discrepancy is the presence of  significant pairwise interactions in Ho${_3}$Mg${_2}$Sb${_3}$O${_{14}}$. Indeed, theoretical work has shown that a strong damping of inelastic excitations is intrinsic to the TIM, and is strongest for $J_{ij} \approx h^x$ \cite{Tommet_1975,Oitmaa_1984,Florencio_1995}. Experiments on model two-singlet systems such as LiTbF$_4$ are qualitatively consistent with this picture \cite{Kotzler_1988,Youngblood_1982,Lloyd_1990}.  Hence, our inelastic neutron-scattering results support a picture of Ho${_3}$Mg${_2}$Sb${_3}$O${_{14}}$ in which frustrated pairwise interactions compete with quantum fluctuations induced by $h^x$.

To obtain a better understanding of our paramagnetic neutron-scattering data, we employ a reciprocal-space mean-field approximation to model the wave-vector dependence of the magnetic diffuse scattering. This approach is exact in the high-temperature limit, and introduces the effect of local spin correlations \emph{via} a reaction-field term that is determined self-consistently \cite{Paddison_2019}. The neutron-scattering intensity is calculated \emph{via} the dynamical susceptibility, which is approximated as
\begin{equation}\label{eq:rpa}
\chi_{\mu}(\mathbf{Q},\omega)=\frac{\chi_{0}(\omega)}{1-\chi_{0}(\omega)[\lambda_{\mu}(\mathbf{Q})-\lambda]},
\end{equation}
where $\chi_{0}(\omega)$ is the single-ion susceptibility obtained by exact diagonalization of $H_\mathrm{SI}$, $\lambda$ is the reaction field \cite{Santos_1980,Paddison_2019}, and $\mu\in \{1,3\}$ labels the normal modes of the Ising system in the same way as for classical mean-field theories \cite{Enjalran_2004}. Full details of this method are given in Ref.~\onlinecite{Paddison_2019} and references therein. We employ this approach to fit the value of $J_\mathrm{nn}$ to the energy-integrated magnetic diffuse scattering. The value of $J_\mathrm{nn}$ is the only free parameter because $h^x=0.85$\,K is fixed (see Section~\ref{sub:cef}). Our fits yield good agreement with the energy-integrated experimental data at $4.2$\,K [Fig.~\ref{fig4}(b)] and at $10$\,K (not shown). However, because damping effects are not included in the mean-field calculation, the energy-resolved response deviates from the experimental results [Fig.~\ref{fig4}(c)]. The qualitative features of our calculations are not strongly sensitive to the value of $J_\mathrm{nn}$, provided that the interactions between pseudo-spins remain frustrated. Nevertheless, $J_\mathrm{nn}=-0.64(4)$\,K yields optimal fits, and we therefore use this value for calculations in the rest of this paper.

We now consider the low-temperature state below $T^*$. In this regime, most of the magnetic neutron-scattering intensity remains diffuse; however, weak magnetic Bragg peaks also appear on top of the magnetic diffuse scattering [Fig.~\ref{fig4}(b)]. The wavevector dependence of the scattering closely resembles observations for the Dy${_3}$Mg${_2}$Sb${_3}$O${_{14}}$, suggesting that a similar spin fragmentation process occurs in Ho${_3}$Mg${_2}$Sb${_3}$O${_{14}}$. In particular, the divergence-full part of a spin-fragmented state describes an ``all in, all out" (AIAO) long-range magnetic ordering involving only a small fraction ($\lesssim \mu/3$) of the total magnetic moment [Fig.~\ref{fig1}(a)]. Rietveld refinements to the weak magnetic Bragg component of our $T = 0.12$\,K data are in good agreement with the AIAO average magnetic structure in Ho${_3}$Mg${_2}$Sb${_3}$O${_{14}}$, consistent with a spin-fragmented state (see Appendix A). Our model calculations, discussed in Section~\ref{sec:THEORY}, provide further evidence that a spin-fragmented state is consistent with the spin Hamiltonian, Eq.~\eqref{eq:EffectiveH}.

However, our measurements also reveal fundamental differences with the CSF state observed in Dy${_3}$Mg${_2}$Sb${_3}$O${_{14}}$. First, the observed magnetic Bragg intensities in Ho${_3}$Mg${_2}$Sb${_3}$O${_{14}}$  are strongly reduced compared to the expected classical value. The magnetic Bragg intensity is proportional to the square of the ordered magnetic moment, which is $\mu/3 =3.3\,\mu_\mathrm{B}$ per site for a CSF state in the absence of chemical disorder [Fig.~\ref{fig1}] \cite{Brooks-Bartlett_2014,Paddison_2016}. In contrast, Rietveld refinements to our $T = 0.12$\,K data indicate an ordered magnetic moment of only $1.70(3)$\,$\mu_\mathrm{B}$ per Ho$^{3+}$ (see Appendix A). 
Importantly, Dy${_3}$Mg${_2}$Sb${_3}$O${_{14}}$ has both a larger ordered moment (2.66(6)\,$\mu_\mathrm{B}$ per Dy$^{3+}$ at $0.20$\,K \cite{Paddison_2016}) and a greater degree of site mixing than Ho${_3}$Mg${_2}$Sb${_3}$O${_{14}}$ (see Section~\ref{sec:EffectHam}). This suggests that chemical disorder cannot fully explain the observed reduction in ordered moment in Ho${_3}$Mg${_2}$Sb${_3}$O${_{14}}$, although it may be a contributing factor. Given the low temperature of our measurement ($\approx$ $0.12$\,K), quantum fluctuations are the most likely alternative explanation.

The second key difference with Dy${_3}$Mg${_2}$Sb${_3}$O${_{14}}$ is that the spin-fragmented state in Ho${_3}$Mg${_2}$Sb${_3}$O${_{14}}$ is accompanied by the persistence of continuous magnetic excitations down to at least $0.12$\,K.  Furthermore, a distinct mode develops at $\omega \approx3$\,K when approaching $T^{\ast}$ and appears clearly separated from the elastic line in $I(Q,\omega)$ below $T^{\ast}$ [red arrows in Fig.~\ref{fig4}(c)]. Above this mode, we observe a high-energy tail extending to $\omega \approx15$\,K that resembles the slow decay from the central peak in the paramagnetic phase. The presence of clear low-temperature spin dynamics over a wide energy range strongly contrasts with canonical classical Ising magnets such as Ho$_2$Ti$_2$O$_7$ \cite{Ehlers_2003} and  Dy${_3}$Mg${_2}$Sb${_3}$O${_{14}}$ \cite{Paddison_2016}, in which the spin dynamics are too slow to observe in neutron-scattering measurements at comparable temperatures \cite{Bai_unpublished}. Meanwhile, spin fragmentation accompanied by inelastic excitations has been observed in Nd$_2$Zr$_2$O$_7$ \cite{Petit_2016, Benton2016quantum}, yet with a key difference that reflects their different physical origins: unlike Nd$_2$Zr$_2$O$_7$ where pinch points appear only in the inelastic channel, the diffuse scattering observed in Ho${_3}$Mg${_2}$Sb${_3}$O${_{14}}$ contains both elastic and inelastic contributions, most of which is elastic within our highest energy resolution of approximately 0.29\,K (25\,$\mu$eV) FWHM on OSIRIS [Fig.~\ref{fig4}(c)]. Overall, the enhancement of inelastic scattering and reduction of the magnetic Bragg intensity provides experimental evidence for quantum excitations above a spin-fragmented ground state in Ho${_3}$Mg${_2}$Sb${_3}$O${_{14}}$.

\section{THEORETICAL MODELING}\label{sec:THEORY}

Our experimental results have revealed two key insights: that the spin Hamiltonian of Ho${_3}$Mg${_2}$Sb${_3}$O${_{14}}$ realizes a frustrated transverse Ising model; and that a spin-fragmented state with quantum spin dynamics exists in Ho${_3}$Mg${_2}$Sb${_3}$O${_{14}}$ at the lowest measurable temperatures. These results identify Ho${_3}$Mg${_2}$Sb${_3}$O${_{14}}$ as an exotic frustrated magnet with unconventional properties. However, two important questions remain. First, how do frustrated interactions, transverse field, and hyperfine coupling conspire to generate the observed quantum spin dynamics? And, second, are these quantum dynamics primarily single-site fluctuations, or do they involve many-body quantum correlations as hypothesized for the QSF state in the introduction?

As a first step towards answering these questions, we perform a theoretical analysis of the spin Hamiltonian, Eq.~\eqref{eq:EffectiveH}. Because this Hamiltonian describes a long-range interacting quantum system with several interactions on the same energy scale, no single theoretical approach can provide a complete description of its properties. We therefore employ two complementary approaches that ultimately reveal how the hyperfine coupling interplays with the transverse field. First, in Section~\ref{sub:ED}, we use exact diagonalization of small clusters to investigate possible quantum correlations between sites. Second, in Section~\ref{sub:MFMC}, we develop a modification of Monte Carlo (MC) simulation in conjunction with real-space mean-field (MF) theory, which enables us to investigate finite-temperature properties on large clusters, for which quantum effects are considered only at the single-site level. 

\subsection{Exact diagonalization} \label{sub:ED}
\begin{figure}[tbp]
	\begin{center}
		\includegraphics[width= 3 in]{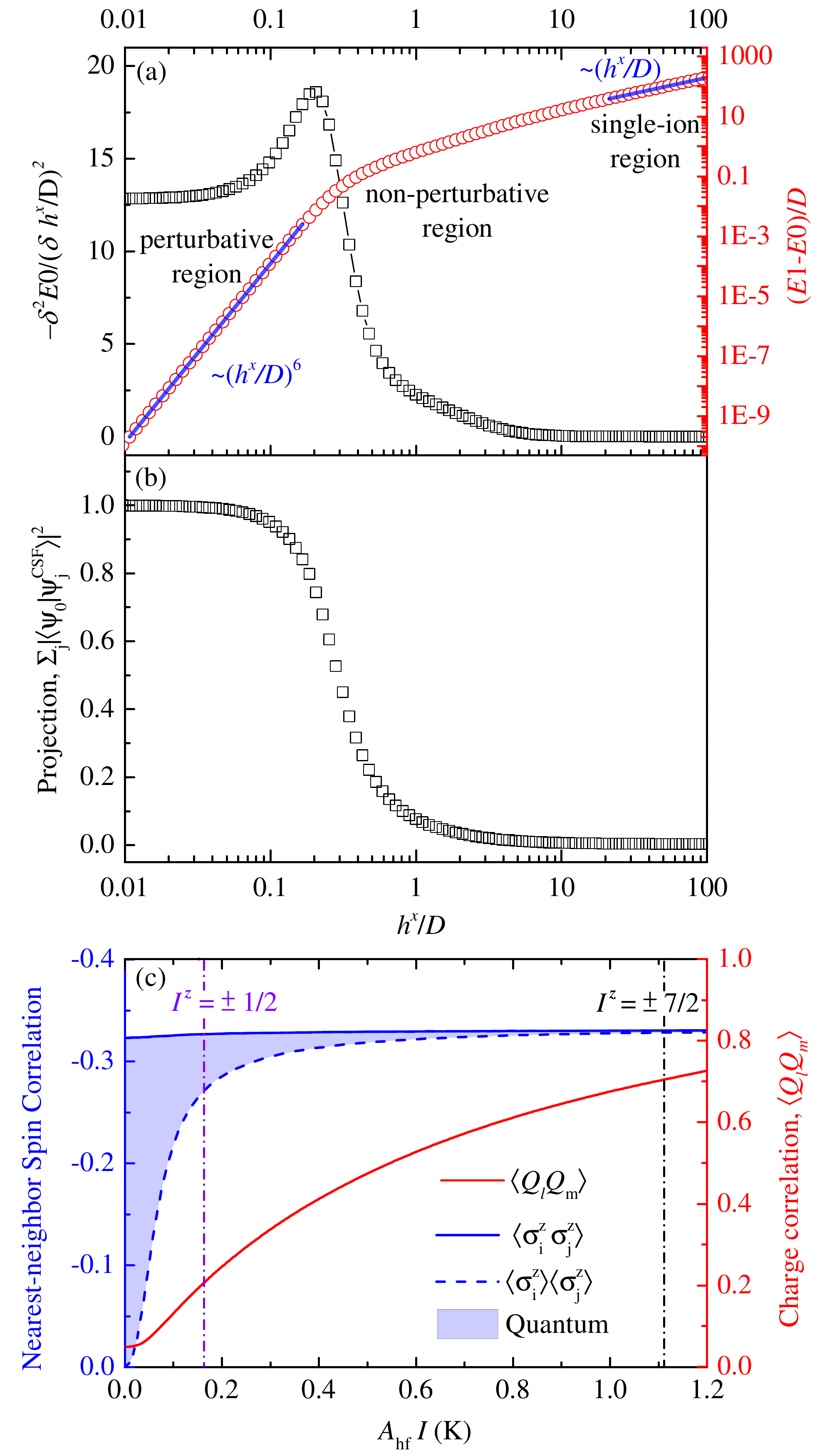}
		\pdfinclusioncopyfonts=1
	\end{center}	\caption{\label{figED}  Results of exact diagonalization of the effective Hamiltonian [Eq.~\eqref{eq:EffectiveH}] on an $N =12$ cluster. (a) and (b) are obtained with $J_{\text{nn}} = 0$ and $A_\textrm{hf} = 0$. (a) Second derivative of the ground-state energy as a function of $h^x/D$ (black squares) and energy gap between the ground state and first excited state (red circles), indicating possible quantum phase transitions. Blue lines represent linear fits to the data. (b) Projection of the ground-state wavefunction into the CSF basis (black squares).  (c) Nearest-neighbor spin correlation (blue) and emergent charge correlation (red) of the ground states as a function of hyperfine coupling strength where $Q_l$ and $Q_m$ are the emergent charges defined as in Fig.~\ref{fig1}. Parameters appropriate for Ho${_3}$Mg${_2}$Sb${ _3}$O${_{14}}$ ($D=1.29$\,K, $J_{\text{nn}} = -0.64$\,K, $h^x = 0.85$\,K) are used in (c).  The two dashed lines indicate nuclear spin sectors with $I^z_i=\pm1/2$ and $\pm7/2$ on each site, respectively.}
\end{figure}

We use exact diagonalization (ED) of small clusters to explore the ground state and low-energy spectrum of the model Hamiltonian, Eq.~\eqref{eq:EffectiveH}. The main results presented here are obtained from diagonalizing a $2\times2$ rhombus-shaped supercell with $N$ = 12 sites; however, our conclusions are robust to the choice of different supercell geometries and the use of 18-site clusters (see Appendix E). 

We start by considering a simple quantum model that contains the essence of Ho${_3}$Mg${_2}$Sb${ _3}$O${_{14}}$: a dipolar kagome ice under a transverse field. In the absence of the hyperfine and exchange couplings, the system is controlled by a single tuning parameter $h^{x}/D$. At the classical point ($h^x/D = 0$), 12 CSF configurations appear to be exactly degenerate for the $2\times2$ cluster. Given that this model is known to order completely in the thermodynamic limit, with a $\sqrt{3}\times\sqrt{3}$ enlargement of the unit-cell \cite{Chern_2011} that matches our rhombus-shaped finite system, these degenerate states are the 6 possible domains of the fully ordered phase plus their time-reversal symmetric counterparts. Distinguishing a CSF phase from this fully-ordered ground-state requires larger size clusters, because the degeneracy of the former grows exponentially with system size while the degeneracy of the later is fixed. 

For the same cluster, an infinitesimal transverse field yields a unique ground state, corresponding to a quantum superposition of the above 12 CSF spin configurations with equal weights, and opens a gap [Fig.~\ref{figED}(a)]. The energy gap to the first excited state scales as $(h^x/D)^6$ until $h^x/D \lessapprox 0.2$ [Fig.~\ref{figED}(a)], reflecting that the leading-order quantum tunneling process between degenerate CSF states corresponds to flipping six spins around a hexagon [Fig.~\ref{fig1}]. Our calculations show that the mixing of states out of the CSF manifold is negligible when $h^{x}/D$ is small, which can be seen by projecting the ground-state wavefunction into the CSF subspace [Fig.~\ref{figED}(b)]. As the transverse field is increased, we see a dramatic decrease of the overlap between the ground state and the CSF manifold at $h^x/D \approx 0.2$, accompanied by a change in the scaling behavior of the energy gap and a peak in the second derivative of the ground-state energy [Fig.~\ref{figED}(a)]. Notably, these features are absent in a transverse Ising model with only nearest-neighbor exchange interactions (see Fig.~\ref{Sfig:SM_ED}), implying possible existence of a quantum phase transition when the long-range dipolar interaction is included in the model. The system eventually crosses over into a paramagnetic phase in the limit $h^{x}\gg D$, corresponding to the limit of isolated single-ion physics.

The above results seem to point to a possible existence of a QSF phase when the transverse field remains a perturbation to the dipolar interaction. However, a recent quantum Monte-Carlo (QMC) work has uncovered that the ground state of a very related system (with spins in plane rather than tilted by 22 degrees) displays a $\sqrt{3}\times\sqrt{3}$ magnetic order up to at least $h^x/D \lessapprox 0.65$~\cite{wang2020tuning}. For such an ordered state, ED will yield a quantum superposition of magnetic domains, so that we cannot distinguish between long-range magnetic order and genuine quantum superposition state without size-scaling studies and consideration of larger size clusters. We also note that even if a quantum superposition state was observed for larger clusters, the CSF subspace should be divided into disconnected topological sectors \cite{Cepas2017colorings} where the composition of the wave-functions depends on the details of the topology. Whether the aforementioned quantum phase transition at finite $h^x$ exists in the thermodynamic limit requires further study.

Notwithstanding these limitations, we can use ED on small clusters to identify the effect of hyperfine coupling as relevant for Ho${_3}$Mg${_2}$Sb${ _3}$O${_{14}}$. We proceed as follows. First, we put Eq.~\eqref{eq:EffectiveH} into a block-diagonal form where each block has a set of fixed nuclear spin numbers $\left\lbrace I_{1}^{z},I_{2}^{z},\cdots,I_{N}^{z}\right\rbrace$ and the hyperfine term is treated as a site-dependent local longitudinal field; this is possible because $I^z$ remains a good quantum number.  Then, we  find the global ground state of the system by diagonalizing the Hamiltonian in each of the 8$^N$ nuclear spin blocks independently. 
The interplay between hyperfine coupling and quantum tunneling is shown in Fig.~\ref{figED}(c). Here, we use the parameters $J_\mathrm{nn}=-0.64$\,K, $h^x=0.85$\,K, and $D=1.29$\,K relevant for Ho${_3}$Mg${_2}$Sb${ _3}$O${_{14}}$, and plot the nearest-neighbor spin and charge correlation functions as a function of the  hyperfine coupling energy scale $A_\mathrm{hf}I$. In the absence of hyperfine coupling, the ground state is in the non-perturbative regime of our ED calculations with strong correlations between spins. 
Although the exact nature of the ground-state in this regime requires further investigation, 
ED remains informative to track the effect of hyperfine interactions. For $A_\mathrm{hf}I=1.11$\,K, as appropriate for Ho${_3}$Mg${_2}$Sb${ _3}$O${_{14}}$, the ground state has $I^z_i=\pm7/2$ for every site. Upon increasing $A_\textrm{hf}$, the nearest-neighbor quantum correlation---defined as the total correlation minus the classical correlation, $\left\langle\sigma_{i}^{z}\sigma_{j}^{z}\right\rangle - \left\langle\sigma_{i}^{z}\right\rangle\left\langle\sigma_{j}^{z}\right\rangle$---diminishes quickly from 100\% at $A_\textrm{hf}I =0$ to only 1\% of the total correlation for $A_\textrm{hf}I = 1.11$\,K [Fig.~\ref{figED}(c)]. This result implies that if quantum correlations are present in the ground state of Ho${_3}$Mg${_2}$Sb${ _3}$O${_{14}}$, they are dramatically suppressed by hyperfine interactions. As the system is thermally excited to hyperfine levels with smaller $\abs{I^z}$, quantum correlations may reappear [Fig.~\ref{figED}(c)]. 

Overall, our ED results suggest that the low-temperature physics of Ho${_3}$Mg${_2}$Sb${ _3}$O${_{14}}$ can be understood in terms of a semiclassical ground-state without quantum correlations between sites and predominantly single-ion quantum fluctuations. However, we speculate that enhanced quantum correlations may appear in the system when other hyperfine states are populated at elevated temperatures.

\subsection{Mean-field Monte Carlo simulations} \label{sub:MFMC}

\begin{figure*}[tbp]
	\begin{center}
		\includegraphics[width= 7 in]{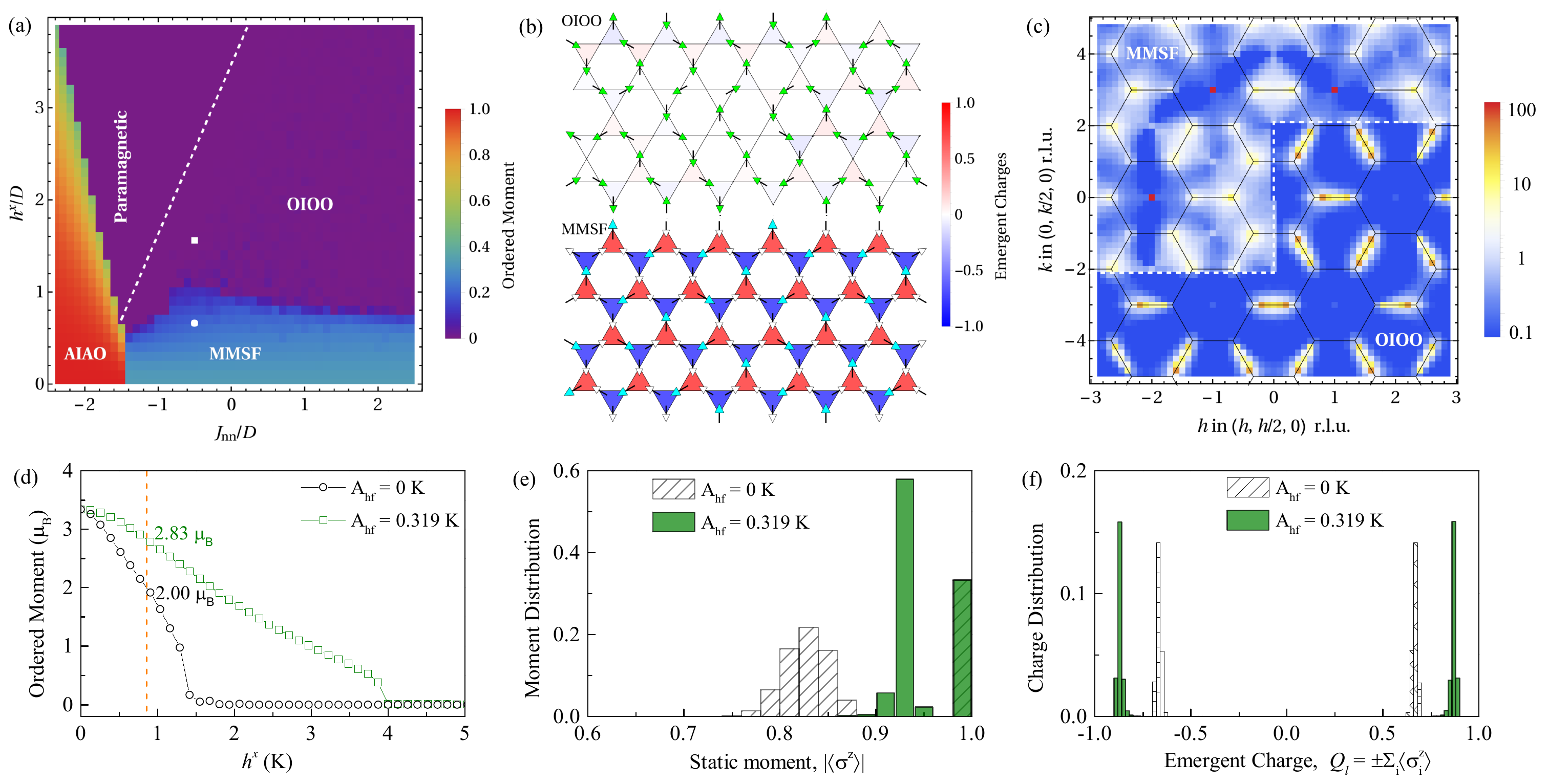}
		\pdfinclusioncopyfonts=1
	\end{center}	\caption{\label{figMF}  Mean field Monte Carlo (MF-MC) simulation results of the mean-field Hamiltonian, Eq.~\eqref{eq:mf}. (a) Phase diagram as a function of $J_{\text{nn}}$/$D$ and $h^{x}$/$D$ with $A_\mathrm{hf} = 0$. The color represents the ordered moment of the all-in-all-out (AIAO) structure at $0.12$\,K. Four distinct phases are observed: AIAO, paramagnetic, one-in-one-out (OIOO), and moment-modulated spin fragmented (MMSF).  The boundary (white dashed line) between the paramagnetic phase and the OIOO phase is obtained from the phase diagram of the static moment (see Fig.~\ref{Sfig:phase}). (b) Representative spin configurations of the MMSF phase  and the OIOO phase, with $h^x$ chosen to be 0.85\,K (white dot in (a)) and 2\,K (white square in (a)), respectively.  The exchange interaction $J_{\text{nn}} = -0.64$\,K and the dipolar interaction $D = 1.29$\,K are the same for all subsequent panels in this figure. The sizes of arrows are scaled with static spins $\langle\sigma^{z}_{i}\rangle$ on each site. Cyan and white arrows denote long ($|\langle\sigma^{z}_{i}\rangle|>0.9$) and short static spins ($|\langle\sigma^{z}_{i}\rangle|<0.9$), respectively. (c) Fourier transform of static spin correlations $\left\langle\left\langle\sigma_i^{z}\right\rangle\left\langle\sigma_j^{z}\right\rangle\right\rangle$ in the MMSF phase (top) and the OIOO phase (bottom). (d) Ordered moment as a function of $h^x$ with and without hyperfine coupling (green squares and black circles, respectively). The orange line corresponds to $h^x =0.85$\,K for Ho${_3}$Mg${_2}$Sb${ _3}$O${_{14}}$.(e) Normalized distribution of static spin lengths, and (f) normalized distribution of magnetic charges for the MMSF phase with and without hyperfine couplings (green and black bars, respectively).}
\end{figure*}

Having established exact results for small spin clusters, we now develop a numerical method capable of approximating local quantum fluctuations in large spin configurations. The essence of our approach is to combine Monte Carlo simulations and exact diagonalization of the mean-field Hamiltonian at each site to obtain equilibrium configurations of static spins. 

We begin by summarizing our approach, which we term ``mean-field Monte Carlo" (MF-MC). 
The mean-field Hamiltonian of a given site $i$ in a spin configuration is given by
\begin{equation} \label{eq:mf}
\mathcal{H}_{i} = h^{z}_i\sigma_i^z + h^{x}\sigma_i^x, 
\end{equation}
where 
\begin{equation}
\quad h^{z}_i = A_\mathrm{hf}I_i^z+\sum_{j}J_{ij}\langle \sigma^{z}_{j}\rangle\, \label{eq:mf_z}
\end{equation}
is the longitudinal field that includes contributions from exchange, dipolar, and hyperfine interactions. Eq.~\eqref{eq:mf} describes a  $2\times 2$ matrix in each sector $n\in\{1,\cdots,8\}$ of the nuclear spin $I_i^{z}$. Diagonalizing these  matrices yields the single-ion eigenstates in the same form as those shown in Fig.~\ref{fig2}(d), \emph{i.e.}, $\ket{\Psi_i}_n = \ket{\psi_i^\pm,I^z_i} $. We obtain the static spin on site $i$ as the diagonal matrix element of $\sigma_{i}^z$,
\begin{align}
\left\langle\sigma^z_{i}\right\rangle = \left\langle \psi_{i}^\pm, I_{i}^{z}| \sigma^z_{i}| \psi_{i}^\pm, I_{i}^{z}\right\rangle = \frac{\pm h^{z}_{i}}{\sqrt{(h^{x})^2+(h^{z}_{i})^2}}.
\end{align}
For zero transverse field, the static spin $\left\langle\sigma^z_{i}\right\rangle$ is a classical Ising variable with unit magnitude on all sites. For nonzero transverse field,  $\left\langle\sigma^z_{i}\right\rangle$ has a magnitude of less than unity, reflecting the nonzero probability of transverse-field-induced spin flips. The probability amplitudes of such spin flips are given by the off-diagonal matrix elements,
\begin{align}\label{eq:off-diagonal}
&\left\langle \psi_{i}^\pm, I_{i}^{z} | \sigma^{z}_{i}| \psi_{i}^\mp, I_{i}^{z}\right\rangle =\frac{-h^{x}}{\sqrt{(h^{x})^2+(h^{z}_{i})^2}}.
\end{align}
To take account of this effect, in our Monte Carlo simulation we do not enforce $\left\langle\sigma^z_{i}\right\rangle = \pm 1$; instead, we compute   $\left\langle\sigma^z_{i}\right\rangle$ on-the-fly from mean-field states $\ket{\psi_{i}^\pm, I_{i}^{z}}$. Our protocol is as follows.
We initialize $\left\langle\sigma^z_{i}\right\rangle$ with random uniformly-distributed values between $-1$ and $1$. One MC step consists of selecting a random site $i$ and proposing an update to the nuclear spin quantum number ($I_{i}^{z}\rightarrow\tilde{I}_{i}^{z}$) and electronic static spin, with the latter selected at random from three possibilities:  (i) maintaining the static spin ($ \left\langle\sigma^z_{i}\right\rangle \rightarrow \left\langle\sigma^z_{i}\right\rangle$), (ii) flipping the static spin ($\left\langle\sigma^z_{i}\right\rangle \rightarrow -\left\langle\sigma^z_{i}\right\rangle$), and (iii) going to one of the two new mean-field states obtained by diagonalizing the mean-field Hamiltonian, Eq.~\eqref{eq:mf}, for which both length and direction of the static spin may be changed ($\left\langle\sigma^z_{i}\right\rangle \rightarrow \left\langle\tilde{\sigma}^z_{i}\right\rangle$). The move is accepted or rejected according to the Metropolis protocol (see Appendix F). Our method is identical to a classical MC simulation with single spin-flip moves in the limit of vanishing transverse field.

As in Section~\ref{sub:ED}, we begin by considering the simple case with $A_{\text{hf}} = 0$. Fig.~\ref{figMF}(a) shows a MF-MC phase diagram as a function of $h^{x}/D$ and $J_{\text{nn}}/D$ at low temperature ($T=0.12$\,K). We observe two trivial phases: first, for dominating $h^{x}$, a paramagnetic ground state with $\langle\sigma^{z}_{i}\rangle=0$ is obtained as anticipated; and, second, for small $h^{x}$ and non-frustrated interactions ($J_\mathrm{nn}/D \ll 0$), a conventional AIAO order is obtained, in which the mean field has the same magnitude on all sites. In contrast, frustrated interactions ($J_\mathrm{nn}/D \gtrsim -1.5$) favor non-trivial states in which the mean field---and hence the static spin---is spatially modulated \cite{Nunez-Regueiro_1992}. Depending on the relative strength of $h^{x}$ compared to the pairwise interactions, this stabilizes two distinct phases. 
For  $h^{x}\gtrsim D$, we find a spin-liquid-like ``one in, one out" (OIOO) phase characterized by a local constraint on every triangle: two out of the three spins are static with roughly equal moments and trace out closed loops in the kagome planes, while the third spin remains entirely dynamic due to the local cancellation of mean fields; hence, emergent magnetic charges are absent [Fig.~\ref{figMF}(b)].
Spin correlations of the OIOO phase give rise to distinct star-like signature on Brillouin zone edges in momentum space, while magnetic Bragg peaks and pinch-point features associated with a spin-fragmented state are absent [Fig.~\ref{figMF}(c)]. 
For  $h^{x}\lesssim D$, we find a phase resembling a CSF state but dressed with a static moment modulation from site to site; we call this phase \textit{moment-modulated  spin fragmented} (MMSF). In this state, quantum fluctuations are manifest for each triangle in the form of one ``long" static spin with a large magnitude, and two ``short" static spins with smaller magnitudes [Fig.~\ref{figMF}(b)]. 
Remarkably, despite a broad distribution of the static spin length [Fig.~\ref{figMF}(e)], sharp magnetic Bragg peaks and diffuse scattering with pinch-point-like features coexist in momentum space [Fig.~\ref{figMF}(c)]. This observation provides a ``smoking gun" for spin fragmentation, which can be understood in the following way. Because each triangle has a ``one in, two out" or ``two in, one out" arrangement of static spins, it supports well-defined emergent magnetic charges, $\left\langle Q_k\right\rangle = \pm\sum_{i\in k}\left\langle \sigma_{i}^{z}\right\rangle$, where $i$ runs over three spins in a triangle $k$. Importantly, $\left\langle Q_k\right\rangle$ form a staggered order with a small variation in magnitude [Figs.~\ref{figMF}(b),(f)]. In the language of spin fragmentation, once we separate the divergence-full part of the MMSF phase (equivalent to the average value of $\left\langle Q_k\right\rangle $ or the ordered moment of the AIAO structure), the remaining part is approximately divergence free and hence gives rise to diffuse scattering with pinch-point singularities. In some sense, the MMSF state is a mean-field mimic of the QSF state where all the quantum many-body correlations are approximated by classical ones.

We now compare our experimental data with MF-MC calculations using the parameters  $D=1.29$\,K, $h^{x} = 0.85$\,K, and $J_{\text{nn}} = -0.64$\,K appropriate for Ho${_3}$Mg${_2}$Sb${ _3}$O${_{14}}$. For $A_\mathrm{hf} =0$, our MF-MC simulations predict a MMSF phase at low temperature [Fig.~\ref{figMF}(a)].
Setting $A_{\text{hf}}$ to the experimental value of $0.319$\,K enhances the classical spin correlations of the MMSF state, and increases the magnitudes of the static moments, ordered moments, and emergent magnetic charges [Fig.~\ref{figMF}(d-f)]. 
Our calculated heat capacity shows a significant improvement over the single-ion calculation and agrees qualitatively with the experimental data [Fig.~\ref{fig3}]. The calculation captures three major features: a broad shoulder between $1$ and $10$\,K, a sharp phase transition into the spin-fragmented phase below $1$\,K, and a large nuclear heat capacity anomaly centered around $0.3$\,K. Our calculated diffuse-scattering patterns agree quantitatively with our energy-integrated neutron-scattering data at $4.2$\,K, and agree qualitatively with the measured momentum dependence of Bragg and diffuse scattering at $0.12$\,K [Fig.~\ref{fig4}(b)]. However, quantitative discrepancies with experiment become increasingly significant as the simulation temperature is lowered: most notably, at $0.12$\,K, the calculation is much sharper than the experimental data and strongly overestimates the intensities of the magnetic Bragg peaks. 
While the MF-MC calculated ordered moment of 2.85\,$\mu_{\text{B}}$ at 0.12\,K is in good agreement with the ED result of 2.80\,$\mu_{\text{B}}$, these values are considerably larger than the experimental result of $1.70(3)\,\mu_{\text{B}}$ [Fig.~\ref{figMF}(d)]. There are two likely explanations for this discrepancy. First, our models neglect the effect of Mg/Ho site disorder, but in fact approximately 3\% of Ho sites are occupied by Mg in Ho${_3}$Mg${_2}$Sb${_3}$O${_{14}}$. Classical calculations suggest that this degree of site mixing can suppress the ordered moment by $\sim$0.5\,$\mu_\mathrm{B}$ \cite{Paddison_2016}; \emph{i.e.}, by a similar amount to quantum fluctuations. 
Second, the large specific-heat anomaly at low temperature makes thermalization of the sample challenging, despite following experimental best practices [Section~\ref{sec:Methods}]. According to our MF-MC calculations, if the true sample temperature is about 85\% of $T^\ast$, the ordered moment will be suppressed to the experimental value  in the absence of site disorder. 

Although our MF-MC calculations cannot describe collective excitations, the density of states of single-ion excitations shown in Fig.~\ref{fig4}(c) provides insight into the dynamics of the MMSF state (see Appendix F). While such excitations are prohibited at low temperature in Dy${_3}$Mg${_2}$Sb${_3}$O${_{14}}$ due to the vanishing off-diagonal neutron dipolar matrix elements, this is no longer the case in Ho${_3}$Mg${_2}$Sb${_3}$O${_{14}}$ due to the presence of the transverse field [Eq.~\eqref{eq:off-diagonal}]. In addition to a large central peak, our calculation shows two small inelastic peaks centered around $\omega=5$\,K and $14$\,K that correspond to flipping short and long static spins, respectively. The relative amount of spectral weight in the elastic \emph{vs.} inelastic channels is qualitatively consistent with the experimental data, but the distinct shoulder in the data appears roughly $3$\,K lower in energy transfer than in the calculation, and the calculation does not reproduce the continuous nature of the excitations. These observations may be a consequence of quantum correlations for the excited states, which are suggested by the ED results [Section~\ref{sub:ED}] but not included in the MF-MC model; hence, quantum calculations of the excitation spectrum would be interesting to explore in future studies.

\section{Discussion \& conclusions}\label{sec:Conclusions}

Our study demonstrates that Ho${_3}$Mg${_2}$Sb${_3}$O${_{14}}$ realizes a frustrated quantum Ising magnet on kagome lattice with three competing energy scales: pairwise interactions between electronic spins, quantum tunneling \textit{via} an intrinsic transverse field, and hyperfine coupling between electronic and nuclear spins. Our experiments uncover overdamped paramagnetic spin dynamics at high temperature and a spin-fragmented state with reduced ordered moment and enhanced single-ion quantum fluctuations at low temperature. The key difference with the CSF phase reported in isostructural Dy${_3}$Mg${_2}$Sb${_3}$O${_{14}}$ is the observable low-energy spin dynamics in Ho${_3}$Mg${_2}$Sb${_3}$O${_{14}}$. These dynamics are a consequence of the intrinsic transverse field emerging from the lower crystallographic symmetry of the Ho$^{3+}$ site in Ho${_3}$Mg${_2}$Sb${_3}$O${_{14}}$ compared to pyrochlore spin ices. Crucially, the transverse field is homogeneous, in contrast to random fields induced by chemical disorder. Therefore, a wide-ranging implication of our study is that symmetry lowering need not be a complicating factor in condensed-matter systems, but can instead enable the observation of simple models of quantum frustration.

Our results motivate further theoretical investigations of the effect of quantum fluctuations on partially ordered states, such as the spin fragmentation observed here for Ho${_3}$Mg${_2}$Sb${_3}$O${_{14}}$. The presence of long-range dipolar interactions and intrinsic transverse field further enriches the problem compared to quantum kagome Ising models studied previously \cite{Moessner_2000, Moessner_2001, Nikolic_2005, Carrasquilla_2015, Wu_2018}.
Our ED calculations on small clusters suggest that the transverse field generates ring-flip tunneling processes that connect degenerate CSF configurations. However, a very recent QMC study on a closely related model shows that this is not sufficient to destroy the low-temperature $\sqrt3\times\sqrt3$ magnetic order, which is seen over a wide range of $h_x/D$ \cite{wang2020tuning}.  Understanding if quantum correlations are present at intermediate temperatures, investigating a possible quantum phase transition with increasing transverse field, and elucidating the role of sub-leading interactions relevant for real compounds, appear as promising directions for future research.

Finally, our results highlight the important role hyperfine interactions can play in frustrated quantum magnets based on rare-earth ions. It is usually assumed that nuclear spins have a spectator role on the behavior of electronic spins, except in the case of singlet electronic ground-states where the hyperfine interaction may induce a cooperative ordering of the combined electronic and nuclear spin system \cite{nicklow_1985,Jensen_1991,Ramirez_1994}. Our results provide a more subtle example compared to this simple picture due to the highly-frustrated nature of spin interactions in Ho${_3}$Mg${_2}$Sb${_3}$O${_{14}}$. Here, hyperfine interactions induce static moments at the single-ion level, and enhance classical correlations at the expense of possible quantum correlations. Yet, due to the frustration, this does not drive the system towards a conventional low-temperature state with complete long-range magnetic order, but instead to a spin-fragmented state that exhibits predominantly single-ion quantum fluctuations. Finally, since all the stable isotopes of trivalent non-Kramers rare-earth ions have non-zero nuclear spin quantum number, hyperfine interactions have significant effects in other frustrated non-Kramers magnets, as for the spin-ice compound Ho$_2$Ti$_2$O$_7$ \cite{Paulsen2019nuclear}, and quantum spin-ice candidates Pr$_2$Zr$_2$O$_7$ \cite{Kimura_2013, Wen_2017} and  Pr$_2$Hf$_2$O$_7$ \cite{Sibille_2018}.

\begin{acknowledgements}
We would like to thank Cristian Batista, Owen Benton, Gia-wei Chern, Yuan Wan, Claudio Castelnovo, Laurent Chapon, Radu Coldea, Si\^{a}n Dutton, Michel Gingras, James Hamp, Peter Holdsworth, Ludovic Jaubert, Gunnar M\"{o}ller, Jeffrey Rau,  Han Yan for helpful discussions, and Allen Scheie for critical reading of the manuscript. We are indebted to Art Ramirez for providing heat-capacity measurements on our s.s. sample. The work of Z.L.D, X.B., J.A.M.P., E.H. and M.M. at Georgia Tech (all analysis, modeling and interpretation work; all synthesis and measurements on s.g. samples) was supported by the U.S. Department of Energy, Office of Science, Office of Basic Energy Sciences Neutron Scattering Program under Award Number DE-SC0018660. The work of Z.L.D and H.D.Z. at the University of Tennessee (all synthesis and measurements on s.s. samples) was supported by the National Science Foundation through award DMR-1350002. H.D.Z acknowledges support from the NHMFL Visiting Scientist Program, which is supported by NSF Cooperative Agreement No. DMR-1157490 and the State of Florida. J.A.M.P. acknowledges support from Churchill College, Cambridge (neutron data reduction), and the U.S.  Department  of  Energy,  Office of  Science,  Basic  Energy  Sciences,  Materials  Sciences  and Engineering  Division (manuscript cowriting). The research at ISIS Neutron and Muon Source was supported by a beam-time allocation from the STFC (U.K.).  The research at Oak Ridge National Laboratory's Spallation Neutron Source and High Flux Isotope Reactor was sponsored by the U.S. DOE, Office of Basic Energy Sciences, Scientific User Facilities Division. Identification of commercial equipment does not imply recommendation or endorsement by NIST.
\end{acknowledgements}

\renewcommand{\thesection}

\section*{Appendix A: Structural and magnetic models}
\begin{figure}[bp]
	\begin{center}
		\includegraphics[width= 3.4 in]{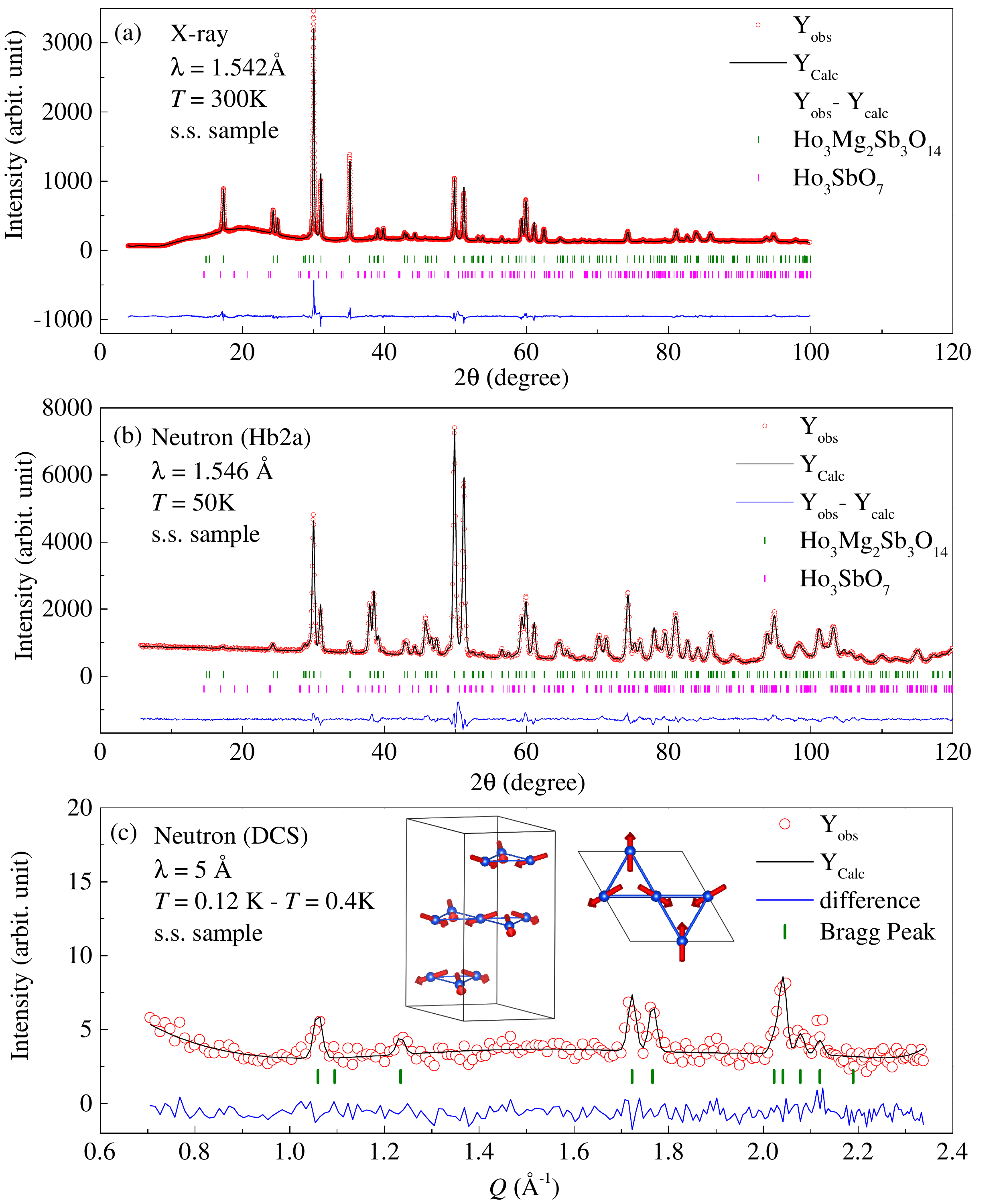}
	\end{center}
	\caption{\label{Sfig:Mag_refine}  Rietveld refinements to diffraction data for the s.s. sample.  Co-refinements of the crystal structure to neutron and X-ray diffraction data are shown in (a) and (b), respectively.  No obvious difference was observed between the X-ray diffraction pattern of s.g. and s.s. samples (not shown here). Refinement of the average magnetic structure to low-temperature magnetic diffraction data is shown in (c). In all panels, experimental data are shown as red circles, Rietveld fits as black lines, and difference (data--fit) as blue lines. Inset: illustrations of the all-in/all-out (AIAO) average magnetic structure within a unit cell and a single kagome layer.}
\end{figure}

The structural model of the s.s. sample of Ho${_3}$Mg${_2}$Sb${_3}$O${_{14}}$ was obtained by Rietveld co-refinements to 50\,K neutron-diffraction data collected using the HB-2A diffractometer [Fig.~\ref{Sfig:Mag_refine}(a)] and 300\,K laboratory X-ray diffraction data [Fig.~\ref{Sfig:Mag_refine}(b)].  Refined values of structural parameters, and selected bond lengths and angles, are given in Table~\ref{tabS1}. 
The canting angle of the Ising axes with respect to the kagome plane is $22.28(2)^{\circ}$ from the co-refinement.  

The average magnetic structure at low temperature was obtained by Rietveld refinement to energy-integrated neutron-scattering data collected on the DCS spectrometer on our s.s. sample [Fig.~\ref{Sfig:Mag_refine}(c)]. We isolated the magnetic Bragg scattering below $T^{\ast}$ by taking the difference between data measured at $0.12$\,K, and $0.4$\,K. The average AIAO magnetic structure belongs to the same irreducible representation as in Dy$_3$Mg$_2$Sb$_3$O$_{14}$, described by $\Gamma_3$ in Kovalev's notation \cite{Kovalev_1993}, which is consistent with a spin-fragmented state \cite{Paddison_2016}. The refined ordered moment is $1.70(3)\,\mu_{\text B}$ per Ho$^{3+}$ with a spin canting angle of 24.9$^\circ$ with respect to the kagome plane. 

\begin{table}[tbp]
	\caption{ \label{tabS1} Crystallographic parameters from Rietveld co-refinement to neutron and X-ray diffraction data of s.s. sample. 
		Anisotropic atomic displacement parameters were used for Mg1. Fixed parameters are denoted by an asterisk ($\ast$). Selected bond lengths and angles are listed.}
	\begin{center}
		\small
		\renewcommand{\arraystretch}{1}%
		\begin{tabular}{cccccc} 
			\hline
			Atom& Site & $x$ & $y$ & $z$ &  Occ. \\
			\hline
			Mg1 & 3a & 0 & 0 & 0 &  1  \\
			Mg2 & 3b & 0 & 0 & 0.5 &  0.905(7)  \\
			Ho(SD) & 3b & 0 & 0 & 0.5 &  0.095(7)  \\
			Ho    & 9d & 0.5 & 0 & 0.5   &   0.968(2)\\
			Mg(SD)& 9d & 0.5 & 0 & 0.5   & 0.032(2)\\
			Sb    & 9e & 0.5 & 0 & 0     & 1  \\
			O1  & 6c & 0 & 0 & 0.1166(4)  &  1    \\
			O2  & 18h & 0.5214(2) & 0.4786(2) & 0.88960(14)  & 1    \\
			O3  & 18h & 0.4694(2) & 0.5306(2) & 0.35556(13)  &  1    \\
			\hline
			&\multicolumn{5}{c}{Neutron diffraction, $T$ = 50 K}\\
			Lattice para. (\AA)  & \multicolumn{5}{c}{$a$ = $b$ = 7.30195(15), \, $c$ = 17.2569(4) } \\
			$B_\mathrm{an.}(\text {Mg1}) ({\text\AA^{2}})$ & \multicolumn{5}{c}{$B_{\text {11}}$  = $B_{\text {22}}$ = 0.0124(22)} \\
			& \multicolumn{5}{c}{$B_{\text {33}}$ = 0.0002(4),\,  $B_{\text {12}}$ = 0.0062(11) } \\
			$B_\mathrm{iso} ({\text\AA^{2}})$ & \multicolumn{5}{c}{$B(\text{Ho})$ = $B(\text{Sb})$ = 0$\ast$} \\
			& \multicolumn{5}{c}{$B(\text{Mg2})$ = 0.07(13),\, $B(\text{O1})$ = 0.14(8)} \\
			& \multicolumn{5}{c}{$B(\text{O2})$ = 0.11(5), \,  $B(\text{O3})$ = 0.24(5)  } \\
			Impurity frac. (\%)	& \multicolumn{5}{c} {$f$(Ho$_3$SbO$_7$) = 2.29(18)}\\
			Bond lengths (\AA) & \multicolumn{5}{c}{  Ho--O1 = 2.278(3)} \\
			& \multicolumn{5}{c}{  Ho--O2 = 2.456(2)} \\
			& \multicolumn{5}{c}{  Ho--O3 = 2.522(3)} \\
			Bond angles ($^{\circ}$) & \multicolumn{5}{c}{ O1--Ho--O2 = 78.69(10)}\\
			& \multicolumn{5}{c}{ O1--Ho--O3 =  76.54(17)}\\
			\hline
			&\multicolumn{5}{c}{X-ray diffraction, $T$ = 300 K}\\
			Lattice para. (\AA) & \multicolumn{5}{c}{$a$ = $b$ = 7.30939(13), \, $c$ = 17.2696(3)} \\
			$B_\mathrm{iso} ({\text\AA^{2}})$ & \multicolumn{5}{c}{Overall  $B$ = 1.38(3)} \\
			Impurity frac. (\%) & \multicolumn{5}{c}{$f$(Ho$_3$SbO$_7$) = 0.75(11)} \\
			Bond lengths (\AA) & \multicolumn{5}{c}{  Ho--O1 = 2.280(3)}\\	
			& \multicolumn{5}{c}{  Ho--O2 = 2.458(2)}\\	
			& \multicolumn{5}{c}{  Ho--O3 = 2.524(3)}\\	
			Bond angles ($^{\circ}$)& \multicolumn{5}{c}{ O1--Ho--O2 = 78.68(10)}\\
			& \multicolumn{5}{c}{ O1--Ho--O3 =  76.55(17)}\\
			\hline
		\end{tabular}
	\end{center}
\end{table}

\section*{Appendix B: Point-charge calculations}

Due to the low point symmetry at the Ho$^{3+}$ site, as many as 15 Stevens operators are required to describe the crystal-field Hamiltonian of the system \cite{walter1984treating}. The number of observables from the inelastic neutron scattering measurements is less than the number of unknown parameters, making conventional fitting procedures impracticable.  To circumvent this problem, we calculated the crystal-field levels and wavefunctions from an effective electrostatic model of point charges using the software package SIMPRE \cite{SIMPRE}. The model considers eight effective oxygen charges surrounding a Ho$^{3+}$ ion, consistent with Rietveld refinements to the powder diffraction data (see Appendix A). The model is then adjusted numerically to match the measured crystal-field spectrum \cite{Dun_2020}. A similar method using a point-charge model has recently been applied to study the crystal-field spectrum in isostructural tripod-kagome compounds with Nd$^{3+}$ and Pr$^{3+}$ ions \cite{Scheie_2018}.
For Ho${3}$Mg${_2}$Sb${_3}$O$_{{14}}$, our point charge model predicts that the two lowest-energy singlets are separated by 1.74\,K ($h^{x}=0.87$\,K), and their wave-functions are given in the total angular momentum basis ($J = 8, J^z = -8,...+8$) by
\begin{align}
    \ket{0} & =  0.690 (\ket{8}+\ket{-8}) - 0.006(\ket{7}-\ket{-7}) \nonumber\\ & - 0.006(\ket{6} + \ket{-6}) - 0.118(\ket{5} - \ket{-5}) \nonumber\\ & + 0.053(\ket{4} + \ket{-4}) + 0.019(\ket{3} - \ket{-3}) \nonumber\\ &  + 0.003(\ket{2} + \ket{-2}) + 0.068(\ket{1} - \ket{-1}) - 0.025 \ket{0}, \nonumber\\
    \ket{1} & = 0.693 (\ket{8}-\ket{-8}) - 0.017(\ket{7}+\ket{-7}) \nonumber\\& + 0.002(\ket{6} - \ket{-6}) - 0.106(\ket{5} + \ket{-5}) \nonumber\\& - 0.008(\ket{4}- \ket{-4}) - 0.065(\ket{3} + \ket{-3}) \nonumber \\ & + 0.003(\ket{2} - \ket{-2}) - 0.048(\ket{1} - \ket{-1}) - 0.000 \ket{0} \nonumber.
\end{align}
The matrix element $\alpha=\langle0|\hat{J}^{z}|1\rangle = 7.79$ gives rise to a total magnetic moment of magnitude 9.74\,$\mu_\textrm{B}$ according to Eq.~\eqref{eq:Moment}.

\section*{Appendix C: Isothermal Magnetization }
\begin{figure}[tbp]
	\begin{center}
		\includegraphics[width= 3.4 in]{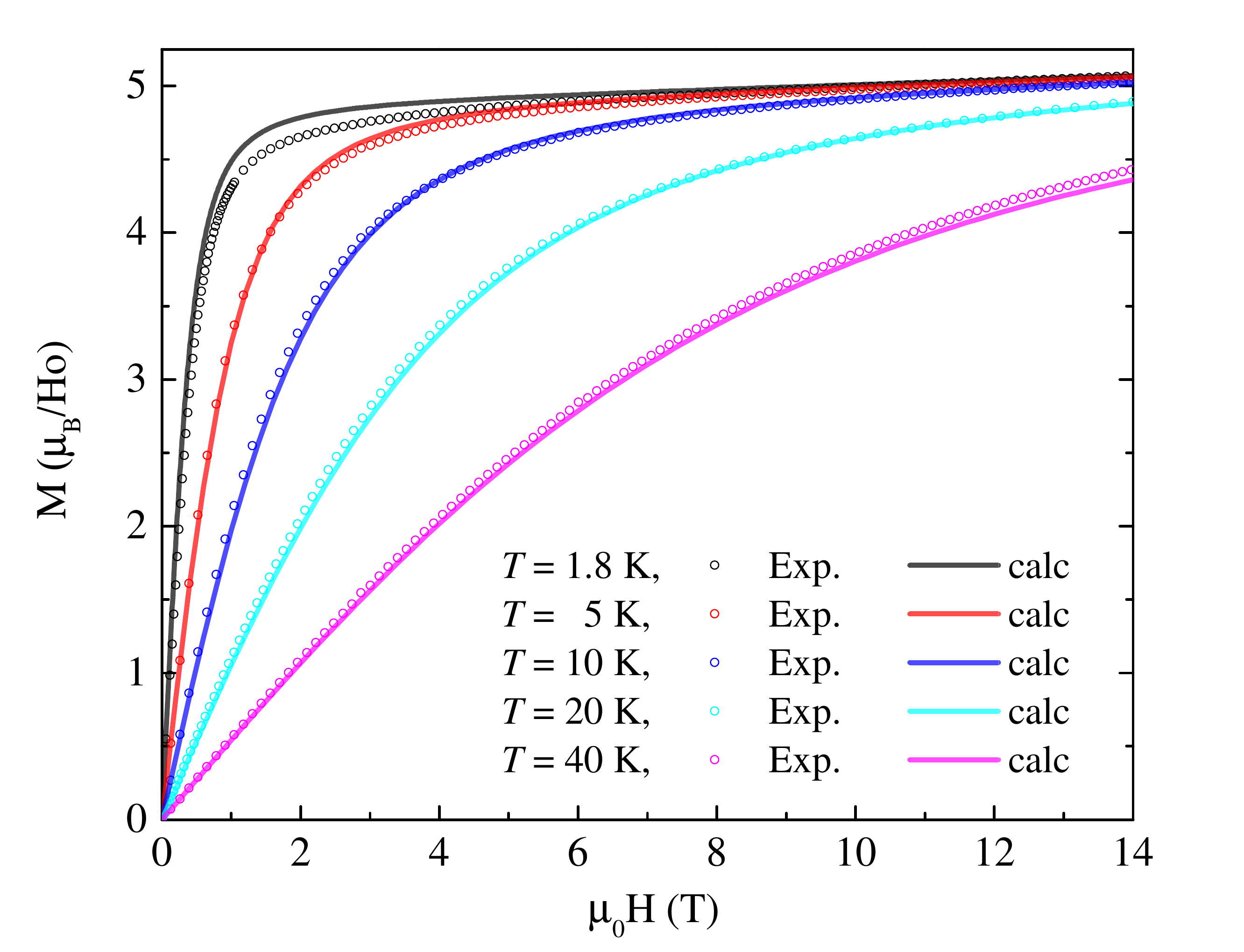}
	\end{center}
	\caption{\label{Sfig:MH}  Isothermal magnetization measurements (open symbols) on the s.g. sample of  Ho${_3}$Mg${_2}$Sb${_3}$O${_{14}}$ at various temperatures. Solid lines were calculated according to Eq.~\eqref{eq:MH_powder} with fixed values of $\alpha = 7.79$, $h_x = 0.85$\,K, $A_\mathrm{hf} = 0.319$ \,K, and a fitted Van-Vleck term $\chi_{vv}/\mu_0$ = 0.0148 $\mu_\textrm{B}/T$. }
\end{figure}

To validate our effective Hamiltonian [Eq.~\eqref{eq:single-ion}], we calculate the isothermal magnetization for Ho${_3}$Mg${_2}$Sb${_3}$O${_{14}}$ and compare it to the experiments. Under an longitudinal magnetic field ${H}$, the effective single-ion Hamiltonian reads,
\begin{equation}
\mathcal{H}  =  h^{x}\sigma_i^x  +  A_\mathrm{hf}I^z\sigma^z + \mu_0\mu_\mathrm{B}\alpha g_J H\sigma^{z}, \label{eq:H_field}
\end{equation} 
With eigenvalues $E_i$ and eigenvectors $\ket{i}$ diagonalized from the above Hamiltonian, the magnetization along the local Ising axis is
\begin{equation}
M^z(H, T)  =  \alpha g_J\sum_{i} e^{-E_i/k_B T} \bra{i}\sigma^z\ket{i} /\sum_{i} e^{-E_i/k_B T}. \label{eq:Mz}
\end{equation}
Along with a Van-Vleck paramagnetic term $\chi_{vv}$, the powder-averaged magnetization is
\begin{equation} 
M_{\text { powder }}(H, T) = \int_{0}^{\pi} M^z(H, T)\sin\theta d\theta + \chi_{vv}H. \label{eq:MH_powder}
\end{equation}

With $\alpha = 7.79$ determined in Appendix~B, $h_x = 0.85$\,K, $A_\mathrm{hf} = 0.319$ \,K, and a fitted value of  $\chi_{vv}/\mu_0$ = 0.0148 $\mu_\textrm{B}/T$, the calculated $M_{\text { powder }}(H, T)$ agrees well with the measured isothermal magnetization in the paramagnetic regime between 1.8\,K and 40\,K [Fig.~\ref{Sfig:MH}]. At low temperatures, such as between 5\,K and 1.8\,K, deviations from the calculation become appreciable due to the development of spin-spin correlations that are not included in this model.

\section*{Appendix D: Paramagnetic effective-field fits}
We used an effective-field approach to calculate the inelastic neutron-scattering
pattern in the paramagnetic phase, based on the Onsager reaction-field
approximation \cite{Santos_1980}.  Full details of this method are given in Ref.~\onlinecite{Paddison_2019}. In this approximation, the
inelastic scattering function is given by
\begin{equation}
S(\mathbf{Q},\omega)=\frac{1}{n\pi[1-e^{-\omega/T}]}\sum_{\mu=1}^{N}\left|\mathbf{F}_{\mu}^{\perp}(\mathbf{Q})\right|^{2}\mathrm{Im}\left[\chi_{\mu}(\mathbf{Q},\omega)\right],\label{eq:s_q_w}
\end{equation}
where $n=3$ is the number of Ho$^{3+}$ ions in the primitive unit
cell, $\omega$ is energy transfer in K, and the susceptibility for
each normal mode $\mu$ is given in Eq.~\eqref{eq:rpa} \cite{Kotzler_1988,Shirane_1971}. The magnetic structure factor is given by
\begin{equation}
\mathbf{F}_{\mu}^{\perp}(\mathbf{Q})=\sum_{i=1}^{N}\mathbf{z}_{i}^{\perp}U_{i\mu}(\mathbf{Q})\exp\left(\mathrm{i}\mathbf{Q}\cdot\mathbf{r}_{i}\right),\label{eq:sf}
\end{equation}
where $\mathbf{Q}$ is the scattering vector, $\mathbf{r}_{i}$ is
the position of magnetic ion $i$ in the primitive cell, and $\mathbf{z}_{i}^{\perp}$
is its local Ising axis projected perpendicular to $\mathbf{Q}$.
The eigenvectors $U_{i\mu}$ and mode energies $\lambda_{\mu}$ are
given at each $\mathbf{Q}$ as the solutions of
\begin{equation}
\lambda_{\mu}(\mathbf{Q})U_{i\mu}(\mathbf{Q})=\sum_{j}J_{ij}(\mathbf{Q})U_{j\mu}(\mathbf{Q}),\label{eq:eigenvalue}
\end{equation}
where the Fourier-transformed interaction $J_{ij}(\mathbf{Q})=\sum_{\mathbf{R}}J_{ij}(\mathbf{R})\exp\left(\mathrm{i}\mathbf{Q}\cdot\mathbf{R}\right)$ includes nearest-neighbor exchange and long-range dipolar contributions,
and $\mathbf{R}$ is the lattice vector connecting atoms $i$ and $j$. The dipolar interaction was calculated using Ewald summation \cite{Enjalran_2004}.
The Onsager
reaction field $\lambda$ is determined by enforcing the total-moment sum rule
\begin{equation}
\frac{1}{nN_{\mathbf{q}}}\sum_{i,\mathbf{q}}\left\langle \sigma_{i}^{z}(\mathbf{q})\sigma_{i}^{z}(-\mathbf{q})\right\rangle =1.\label{eq:onsager_rpa-1}
\end{equation}

The scattering intensities were calculated as
\begin{equation}
I(\omega)=C\left[\dfrac{\mu f(Q)}{\mu_{\text{B}}}\right]^2\int_{Q_{0}}^{Q_{1}}\left\langle S(\mathbf{Q},\omega)\right\rangle _{Q}\mathrm{d}Q,\label{eq:intensity_edep}
\end{equation}
where $Q_{0}=0.4$\,\AA$^{-1}$ and $Q_{1}=1.6$\,\AA$^{-1}$, and
\begin{equation}
I(Q)=C\left[\dfrac{\mu f(Q)}{\mu_{\text{B}}}\right]^2\int_{-\omega^{\prime}}^{\omega^{\prime}}\left\langle S(\mathbf{Q},\omega)\right\rangle _{\omega}\mathrm{d}\omega,\label{eq:intensity_qdep}
\end{equation}
where $\omega^{\prime}=20$\,K, angle brackets here denote numerical spherical
averaging, $f(Q)$ is the Ho$^{3+}$ magnetic form factor \cite{Brown_2004}, $\mu=10\,\mu_\mathrm{B}$ is the total magnetic moment per Ho$^{3+}$, and the constant $C=\left(\gamma_\mathrm{n} r_\mathrm{e}/2\right)^2=0.07265 $\,barn. The integrals were performed numerically.

\section*{Appendix E: Exact diagonalization} 
\begin{figure}[tbp]
	\begin{center}
		\includegraphics[width= 3.4 in]{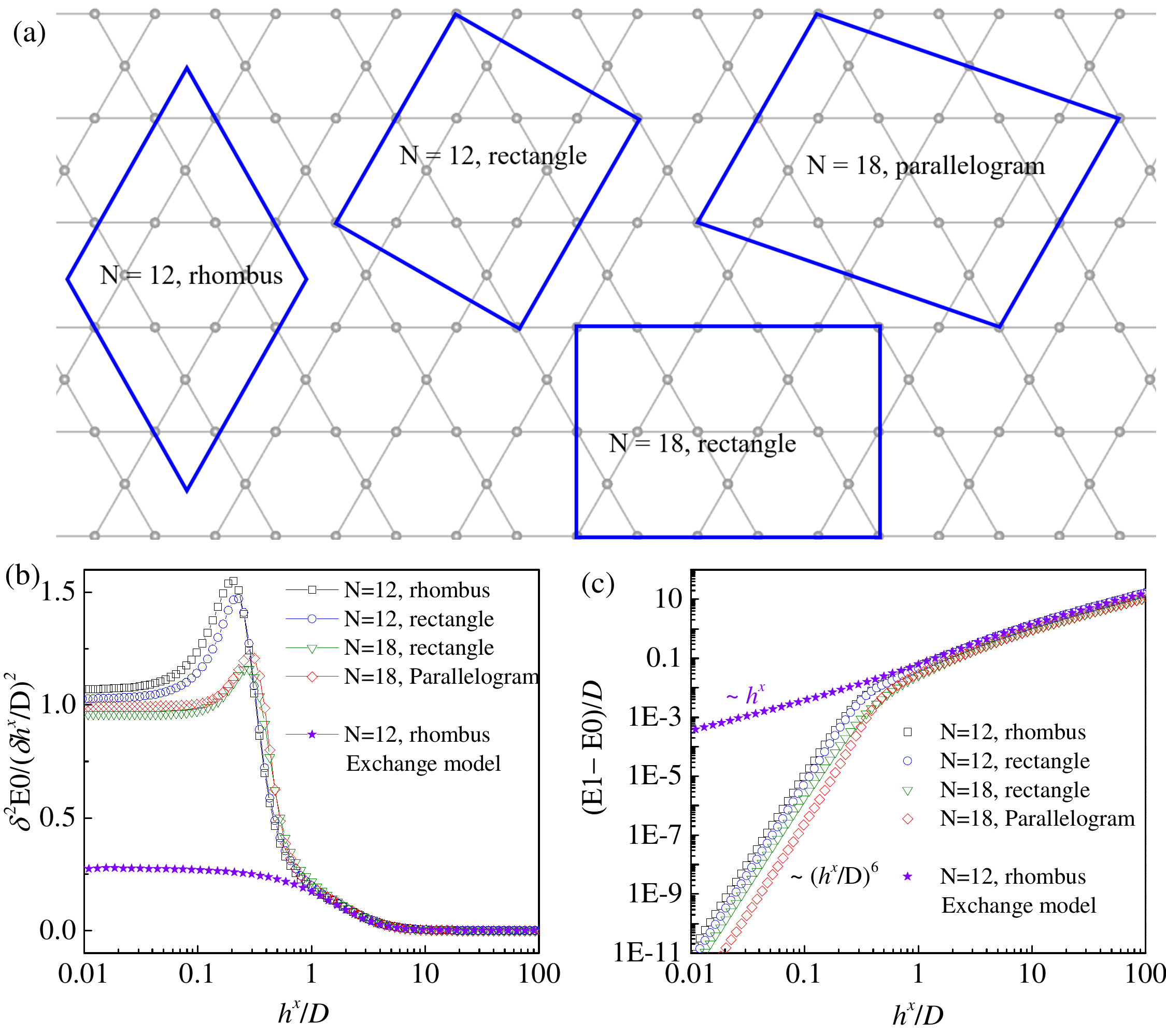}
	\end{center}
	\caption{\label{Sfig:SM_ED}  (a) Different clusters used for the exact diagonalization (ED) calculations of size $N$ = 12, 18.  ED results of Eq.\,\ref{eq:EffectiveH} with $J_{\text{nn}}$ = 0, and $A_\textrm{hf}$ = 0, showing (b) second derivative of the ground state energy, and (c) energy gap between the ground state and first excited state.   ``Exchange model" denotes a transverse Ising model with nearest neighbor exchange couplings only.}
\end{figure}

Due to computational limitations, our ED calculations were restricted to one tripod kagome layer. Cluster sizes of $N = 12$ and $18$ with different shapes were studied under periodic boundary conditions.  Long-range dipolar interactions were summed over periodic copies of the cluster cells up to a distance of 500\,$r_\mathrm{nn}$.

The existence of a quantum phase transition at small $h^x/D$ as well as the the $(h^x/D)^6$ dependence of the exciton energy are independent of size or shape of the clusters [Fig.~\ref{Sfig:SM_ED}]. When truncating the dipolar interaction at the first nearest neighbor, our dipolar TIM becomes an exchange TIM model that has been investigated previously \cite{Moessner_2001,Moessner_2001,Nikolic_2005}. Our results are consistent with these studies for which the ground states obtained at low field are continuously connected to the high-field paramagnetic state with a $h^x/D$ dependence of the exciton energy.

When considering the hyperfine interactions, Eq.~\eqref{eq:EffectiveH} can be put into a block-diagonal form because $I^z$ remains a good quantum number. Then each block has a set of fixed nuclear spin numbers $\left\lbrace I_{1}^{z},I_{2}^{z},\cdots,I_{N}^{z}\right\rbrace$ and the hyperfine term is treated as a site-dependent local longitudinal field.  Therefore, we find the global ground state by diagonalizing the Hamiltonian in each of the 8$^N$ nuclear spin blocks independently. 
As expected, the ground states of the Hamiltonian lives in the nuclear spin blocks where $I^z_i=\pm7/2$ for every site.
The nuclear spin configurations of these ground states match with those electronic spins of the CSF states and thus possesses with the same degeneracy. In other words, the degeneracy of a classical Ising system is resorted in a quantum system through the nuclear spin channel. Recall that the ground state in the absence of hyperfine interaction is a superposition of all CSF basis states. The key effect of the hyperfine interaction is to promote the probability weight of one of the basis states over the others, which means it is suppressing quantum effects and driving the system to the classical limit, as discussed in the main text.

\section*{Appendix F: Mean-field Monte Carlo simulations} 
 
\begin{figure}[tp]
	\begin{center}
		\includegraphics[width= 0.45\textwidth]{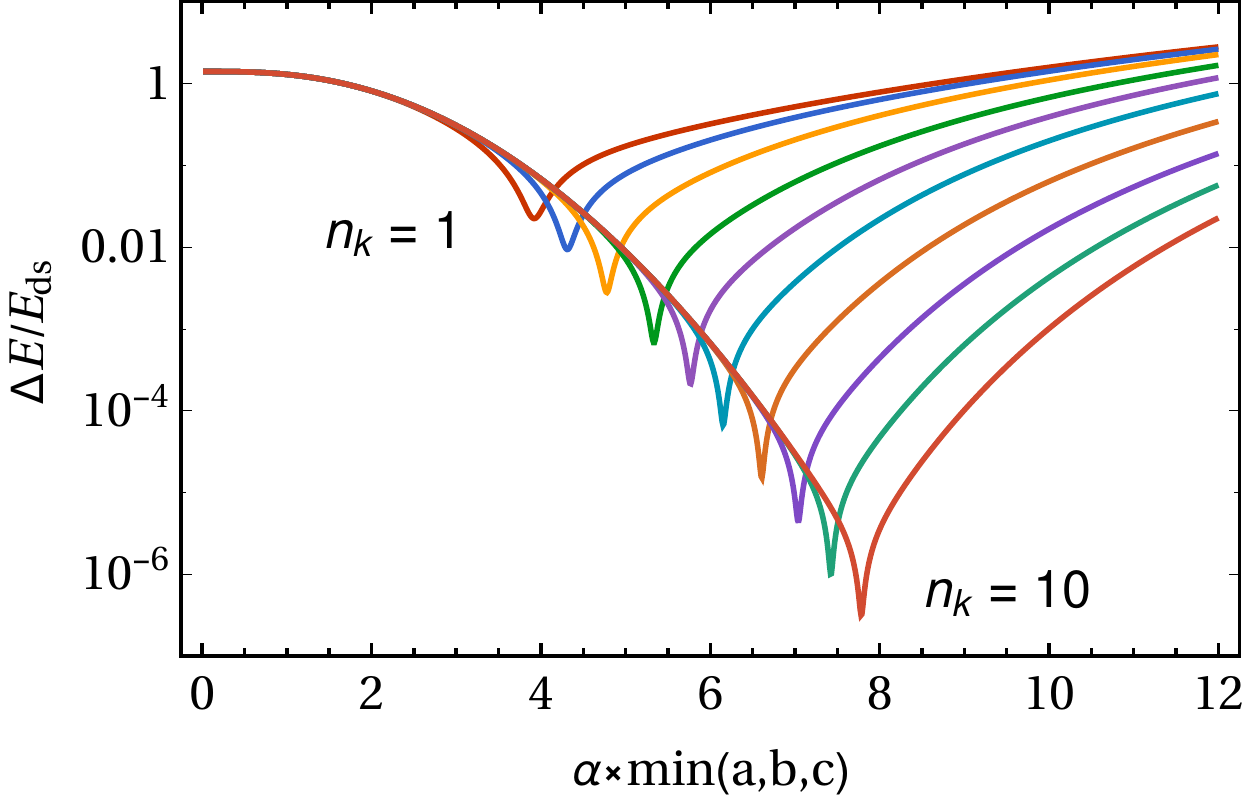}
	\end{center}
	\caption{\label{Sfig:SM_Ewald}  The relative difference of total energy between the Ewald summation and direct summation as a function of the splitting parameter $\alpha$ and reciprocal space cutoff $n_k$. The real space cutoff is taken to be $r_{\text{c}}=\min(a,b,c)/2$, where $a,b$ and $c$ are the dimension of  $5\times3\times2$ orthorhombic supercell. The real space cutoff in the direct sum is $r_{\text{c}}=1000r_{\text{nn}}$. The calculation is averaged over 50 random spin configurations. We take the reciprocal space cutoff $n_k=10$ and the best splitting parameter $\alpha = 7.78/\min(a,b,c)$. The same procedure is performed to find the best splitting parameter $\alpha = 8.10/\min(a,b,c)$ for the $10\times6\times4$ simulation box with the same reciprocal-space cutoff.}
\end{figure}

\begin{figure*}[tbp]
	\begin{center}
		\includegraphics[width= 7 in]{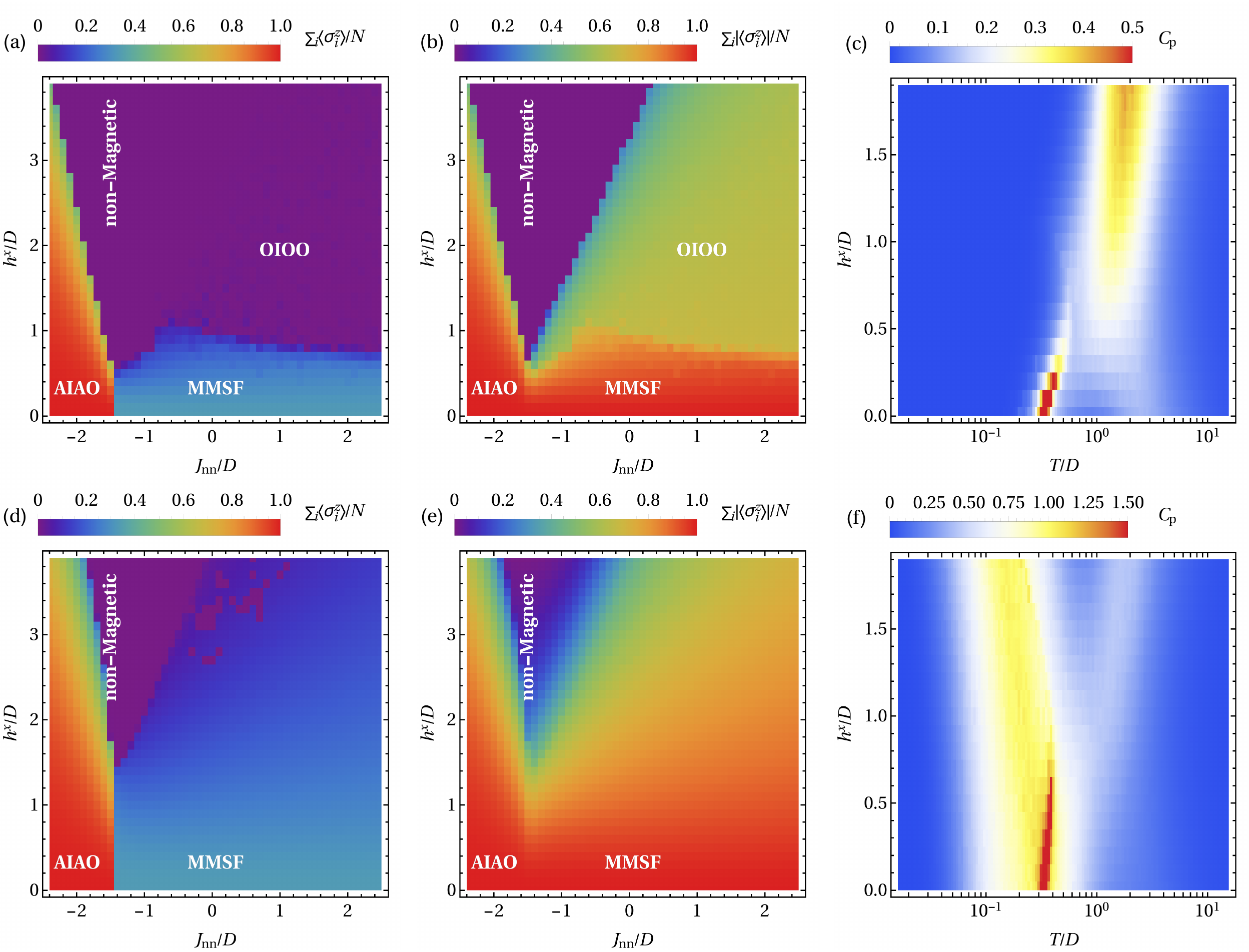}
		\pdfinclusioncopyfonts=1
	\end{center}	\caption{\label{Sfig:phase} Comparison of the ordered moment (a,d), the static moment (b,e) and the specific heat (c,f) without the hyperfine coupling (top three panels), and with the hyperfine coupling  $A_{\text{hf}} = 0.319$\,K (bottom three panels) from our MF-MC simulations. The dipolar interaction strength is $1.29$\,K. The temperature of (a), (b), (d) and (e) is  $0.12$\,K. The exchange interaction strength in (c) and (f) is $-0.64$\,K. The OIOO phase is destroyed by including the hyperfine coupling as shown by comparing (a) and (d).}
\end{figure*}

The key idea of our mean-field Monte Carlo simulation is to generalize the local states from the eigenstates of $\sigma^z$ to those of the single-ion mean-field Hamiltonian. These two sets of eigenstates are identical at the limit of $h^x = 0$. The primary effect of the transverse field in our current calculation is to introduce quantum fluctuation at the level of a single site. 

We keep track of the expectation values of $\sigma^z_i$, $\sigma_i^x$ and $I^z_i$ at each site during the simulation. The energy of the system is computed by replacing the spin operators in the Hamiltonian Eq.~\eqref{eq:EffectiveH} by their corresponding expectation values. The updates of static spin $\langle \sigma^z_i \rangle$ described in the main text are accompanied by corresponding changes in $\langle \sigma^x_i \rangle$; \emph{e.g.}, $\langle \sigma^x_i \rangle$ becomes $-\langle \sigma^x_i \rangle$ if the static spin is flipped. When a new mean-field state is proposed, we update both $\langle \sigma^x_i \rangle$ and $\langle \sigma^z_i \rangle$. 

We used an orthorhombic $5\times3\times2$ supercell of the crystallographic unit cell (see Appendix A) containing $N=540$ sites in six kagome layers to obtain the heat capacity and the phase diagram shown in Figs.~\ref{fig3} and \ref{figMF}, respectively.  A $10\times 6 \times 4$ simulation box containing $N=4320$ sites was used to compute the Fourier transform of the static spin correlation function shown in Fig.~\ref{figMF}(c), and the powder neutron-scattering patterns shown in Fig.~\ref{fig4}. The long-range dipolar interaction was treated by Ewald summation with tinfoil boundary condition at infinity \cite{Leeuw_1980,Melko_2004}. The interaction matrix was computed only once at the beginning of the simulation using the formulae for non-cubic unit cells given in Ref.~\onlinecite{Aguado_2003}. Suitable Ewald parameters were chosen by comparing with the result from direct summmation [Fig.~\ref{Sfig:SM_Ewald}]. The canted local $\textbf{z}$-axis of $22.28^{\circ}$ obtained from the structure refinement was implemented. A simulated annealing process was performed during simulations. We started with random spin configurations at $20$\,K and cooled the system to $0.1$\,K with an exponential rate of $0.95$. At each temperature, we performed $10000$ equilibration sweeps, then collected samples after every 30 sweeps. The heat capacity shown in Fig.~\ref{fig3} was averaged over 30000 samples. One sweep consisted of $N$ MC steps, where $N$ is the number of sites in the simulation box. 
 
The calculated magnetic scattering shown in Fig.~\ref{fig4}(b) was obtained as the sum of static diffuse $I_{\text{diff}}(Q)$, Bragg $I_{\text{Bragg}}(Q)$, and inelastic $I_{\text{inelastic}}(Q)$ contributions, minus the high-temperature paramagnetic $I_{\text{para}}(Q)$ contribution,
\begin{align}
	I_\text{sub}(Q)= I_{\text{diff}}(Q)+I_{\text{Bragg}}(Q)+I_{\text{inelastic}}(Q)-I_{\text{para}}(Q),
\end{align}
where the Bragg and diffuse contributions are calculated following Ref.~\onlinecite{Paddison_2016}.
The inelastic contribution was given by
\begin{align}
	I_{\text{inelastic}}(Q) = \dfrac{2}{3}C\left[\dfrac{\mu f(Q)}{\mu_{\mathrm{B}}}\right]^2\dfrac{1}{N}\sum_{i}\left(1-|\langle\sigma_{i}^{z}\rangle|^2\right).
\end{align}
The $Q$-averaged inelastic spectrum $I(\omega)$ was calculated as
\begin{align}
I(\omega) = \dfrac{2}{3}C\left[\dfrac{\mu}{\mu_{\mathrm{B}}}\right]^2\dfrac{1}{N}\sum_{i}\left( W_i \delta(\omega)+(1-W_i)\delta(\omega-E_{i})\right),
\end{align}
where $E_i = 2\sqrt{(h_{i}^{x})^{2}+(h_{i}^{z})^{2}}$ is the energy of the single-spin excitation and 
$W_i = (h_{i}^{z})^{2}/((h_{i}^{x})^{2}+(h_{i}^{z})^{2})$ is
the elastic spectral weight. The green line shown in Fig.~\ref{fig4}(c) was obtained by convoluting $I(\omega)$ with the experimental energy resolution at the elastic line (FWHM $\approx$ $1$\,K).

\bibliography{HoTKL}

\end{document}